\documentclass{article}

\usepackage{arxiv}

\usepackage[utf8]{inputenc} % allow utf-8 input
\usepackage[T1]{fontenc}    % use 8-bit T1 fonts
\usepackage{nicefrac}       % compact symbols for 1/2, etc.
\usepackage{microtype}      % microtypography

\usepackage{hyperref}       % hyperlinks
\usepackage{url}            % simple URL typesetting
\hypersetup{colorlinks,allcolors=black}

\usepackage{amsmath}
\newcommand\numberthis{\addtocounter{equation}{1}\tag{\theequation}}
\usepackage{amsfonts}
\usepackage{amssymb}

\usepackage{physics}
\usepackage{mathtools}

\newcommand\z[2][\rlap]{\ifx.#2\mperiod[#1]\else\ifx,#2\mcomma[#1]\fi\fi}
\newcommand\mperiod[1][\rlap]{#1{\;\;\;.}}
\newcommand\mcomma[1][\rlap]{#1{\;\;\;,}}

\usepackage{multirow}
\usepackage{array}
\usepackage{makecell}
\newcolumntype{x}[1]{>{\centering\arraybackslash}p{#1}}

\usepackage{tikz}

\graphicspath{ {/Users/jantinebroek/Documents/Pictures/} }
\usepackage{graphicx}
\usepackage{subcaption}
\usepackage{float}
\usepackage[font={small}]{caption}
\usepackage{pdflscape}
\usepackage{afterpage}
\usepackage{flafter}
\usepackage{fancyhdr}
\fancypagestyle{lscape}{% 
	\fancyhf{} % clear all header and footer fields 
	\fancyfoot[LE]{}
	\fancyfoot[LO] {}
	 
	}

\usepackage{xcolor}
\usepackage{hhline}

\usepackage{wrapfig}
\usepackage{tcolorbox}

\title{Generalisation of neuronal excitability allows for the identification of an excitability change parameter that links to an experimentally measurable value.}

\author{
  Jantine A.C.~Broek\\
  Department of Electrical Engineering and Computer Science\\
  University of Liege\\
  Liege, Belgium \\
  \texttt{abroek@uliege.be} \\
   \And
  Guillaume~Drion\\
  Department of Electrical Engineering and Computer Science\\
  University of Liege\\
  Liege, Belgium \\
  \texttt{gdrion@uliege.be} \\
}

\begin{document}
\maketitle

\begin{abstract}
	Neuronal excitability is the phenomena that describes action potential generation due to a stimulus input. Commonly, neuronal excitability is divided into two classes: Type I and Type II, both having different properties that affect information processing, such as thresholding and gain scaling. These properties can be mathematically studied using generalised phenomenological models, such as the Fitzhugh-Nagumo (FHN) model and the mirrored FHN (mFHN). The FHN model shows that each excitability type corresponds to one specific type of bifurcation in the phase plane: Type I underlies a saddle-node on invariant cycle (SNIC) bifurcation, and Type II a Hopf bifurcation. However, the difficulty of modelling Type I excitability is that it is not only represented by its underlying bifurcation, but also should be able to generate frequency while maintaining a small depolarising current. Using the mFHN model, we show that this situation is possible without modifying the phase portrait, due to the incorporation of a slow regenerative variable. We show that in the singular limit of the mFHN model, the time-scale separation can be chosen such that there is a configuration of a classical phase portrait that allows for SNIC bifurcation, zero-frequency onset (due to the infinite period) and a depolarising current, such as observed in Type I excitability. Using the definition of slow conductance, $g_s$, we show that these mathematical findings for excitability change are translatable to reduced conductance based models and also relates to an experimentally measurable quantity. This not only allows for a measure of excitability change, but also relates the mathematical parameters that indicate a physiological Type I excitability to parameters that can be tuned during experiments. Therefore, not only did we indicate a region in a generalised phenomenological model that translates to the physiological conditions of Type I and Type II excitability, but also found mathematical parameter that are measurable in an experimental setting.
\end{abstract}

% keywords can be removed

\section{Introduction}
Excitability is a key feature of neurons in which the qualitative response characteristic is critically dependent on features of different types of ion channels. To capture neuronal excitability in a mathematical framework, canonical models use dynamical analysis to distinguish the class of excitability solely based on their underlying bifurcation. In the phase plane, Type I is typically classified by a saddle-node on invariant cycle (SNIC) bifurcation and Type II by a subcritical Andronov-Hopf bifurcation. However, this is an incomplete story, as other features in the phase plane do not comply with the biological observations and mathematical formulation of Type I excitability. In this work we show that the classification of Type I and Type II excitability solely based on their underlying bifurcation can be further generalised by taking into account the additional constraints for Type I excitability, which are, besides SNIC bifurcation, a zero-frequency onset due to the infinite period cycle, and a small transmembrane depolarising current during the inter-spike interval (ISI). The incorporation of a slow regenerative parameter, such as done in the mirrored Fitzhugh-Nagumo (mFHN) model, allows for the existence of a region that meets the all the requirements for Type I excitability, i.e. SNIC bifurcation, zero-frequency onset and a small transmembrane depolarising current during ISI. This finding is not only present in phenomenological models, but is also found in conductance based models (CBMs) and can be translated to a experimentally measurable value. By translating the generalisation of neuronal dynamical behaviour to different types of models and actual measurable experimental values allows for the identification of parameters that are not only descriptive but contain predictive properties for neuronal behaviour.  \par

An excitable system is able to switch between a resting state and the ability to elicit spiking behaviour. The dynamics of neuronal excitable are phenomenologically modelled by the Fitzhugh-Nagumo (FHN) model \cite{fitzhugh1955mathematical}. The classical FHN is well known, and is a two dimensional reduction and simplification of the four dimensional Hodgkin-Huxley model \cite{hodgkin1948local, hodgkin1952quantitative}. This mathematical reduction allows the study of neuronal dynamics using phase plane analysis. Hodgkin recognised three different classes of neuronal excitability, `Type I', `Type II' and `Type III', using their experimentally obtained frequency-current (FI) curve. In this study we focus on Type I and Type II, as the FI curve for Type III is undefined. The excitability classes underlie differences in initial firing rate and spike initiation, i.e. a local difference in state change. Type I neurons, such as regular spiking pyramidal cells \cite{tateno2004threshold}, have the property of continuous gain scaling in the FI curve and is therefore able to generate low firing frequencies depending on the strength of the applied current \cite{ermentrout1996type, rinzel1989analysis, connor1971prediction, connor1977neural}. In theory, this means that the the period between two spikes, i.e. the ISI, is able to go to infinity. This underlies one of the constraints of Type I excitability, which is zero-frequency onset as a result of the infinite period cycle. In order to meet the Type I property of continuous gain scaling, a mathematical model must be able to allow for a situation in which the constraint of zero-frequency onset is met while the system undergoes SNIC bifurcation. Another biological observation during Type I excitability, is the existence of an extremely small transmembrane depolarising current during the ISI \cite{khaliq2008dynamic}. This small transmembrane current is necessary to maintain a continuous firing at a low frequency rate. Therefore, to model gain scaling during Type I excitability, the phase plane should reflect the biological observations that typifies Type I excitability and the mathematical formulation of them. Fast-spiking inhibitory interneurons \cite{tateno2004threshold} and brainstem mesenchymal V cells are Type II and contain a non-continuous FI curve, which abruptly moves from quiescence to fast spiking \cite{rinzel1989analysis, izhikevich2007dynamical, ermentrout2010mathematical}. As the properties of Type I (gain scaling) and Type II (thresholding) are different, they have consequences for input processing and computation \cite{herz2006modeling, koch1997computation, schwartz2001natural, ratte2013impact, brenner2000adaptive}. \par

%time-scale separation
The excitability change neurons relates to conductance changes in the post-synaptic membrane. In mathematical models, this is generalised in two-dimensional models by exploiting the separation of time-scales and its effect on spike initiation. The time-scale coupling factor, $\epsilon = \frac{1}{\tau_w}$ where $\tau_w >> 1$ (see eqn. \ref{eq:FHN_eqn} and eqn. \ref{eq:mFHN_eqn}), reflects the accepted strong separation between the fast voltage dynamics and the sodium ion channel activation kinetics, and the remaining kinetics of the slow gating ion channels \cite{gerstner2014neuronal}. By changing $\epsilon$, the neural excitability system changes state via either a SNIC or a Hopf bifurcation, where the critical point of bifurcation transition is $\epsilon_{crit}$. Changing $\epsilon$ does not change the nullclines, however, the dynamics of the system are changed. Bifurcation theory relates these mathematical signatures to the physiological classification of the neuron. Specifically, we are interested in the situation where the value of $\epsilon$ allows for two of the three constraints of Type I excitability, which is a SNIC bifurcation and an zero-frequency onset in the system. \par

\begin{figure}
	\centering
	\includegraphics[width=0.8\linewidth]{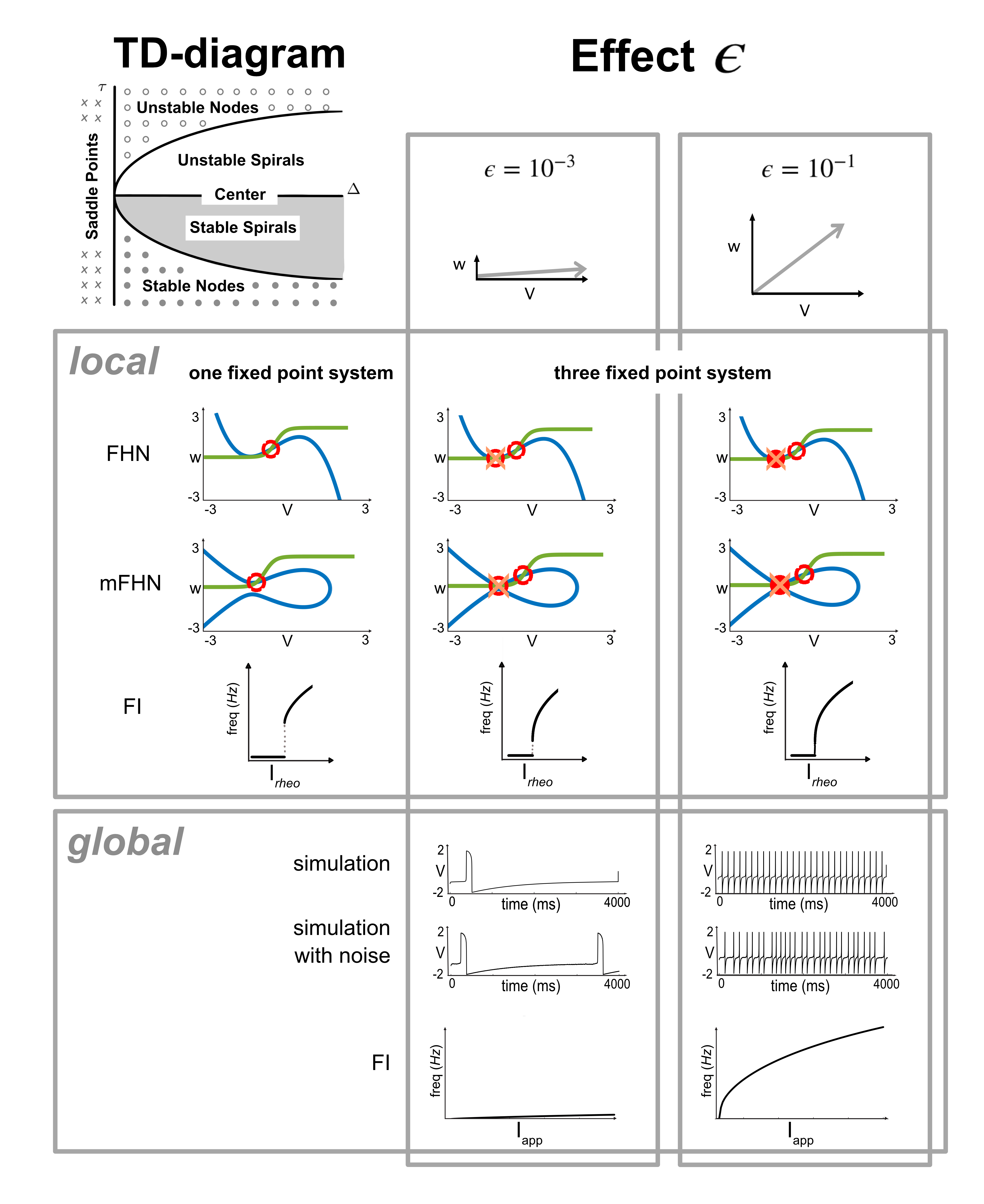}
	\caption{\textbf{The local and global effects of $\epsilon$ in the classical Fitzhugh-Nagumo (FHN) model and the mirrored Fitzhugh-Nagumo (mFHN) model.} \textbf{TD-diagram} is used for the classification of the fixed point stability, as described in \cite{strogatz2018nonlinear} (pg. 138). Below the TD-diagram, the \textit{local} effect and \textit{global} effect in the phase planes are shown for a system with one fixed point (left) and three fixed points (middle and right), for both the classical FHN and mFHN model. In the system with one fixed point, the fixed point is destabilised via a subscritical Andronov-Hopf bifurcation, reflecting Type II excitability. This type of excitability is reflected as a jump at the rheobase, $I_{rheo}$, in the frequency-current (FI) curve. In a system with three fixed points, the \textbf{effect of $\epsilon$} is crucial in both the local (upper part) and global (lower part) excitability signatures. The time-scale coupling factors is defined as $1/\tau_w = \epsilon$. The effect of a large time-scale separation ($\epsilon = 10^{-3}$) and a small time-scale separation are visualised ($\epsilon = 10^{-1}$) for a system with three fixed points. When the time-scale coupling, $\tau_w$ is large, the $\epsilon$-value is small, and the vector field is almost horizontal. On the other hand, a small $\tau_w$-value results in a large $\epsilon$-value and an oblique vector field. In a system with three fixed points, the effect of $\epsilon$ on the bifurcation type results in two possibilities: either the left fixed point is destabilised (middle inset, left open circle) via a subcritical Hopf bifurcation before collision with the saddle (cross) or the left fixed point remains stable (closed circle) and collides with the saddle via SNIC bifurcation (right insets). The FI curves show that the jump from rest to spiking at $I_{rheo}$ is small for Type II excitability. At $I_{rheo}$ in Type I excitability, the FI curve is continuous, although the lowest frequencies are hard to measure both computationally and experimentally. Theoretically, Type I excitability should reach an infinite period. The \textit{global} effect of $\epsilon$ shows that overall the frequencies decreases when $\epsilon$ is small. This is both seen in the simulations and the global effect on the FI curve. Adding noise results in either a delay in spiking or a faster spike, however, low $\epsilon$-value are less affected by noise compared to high $\epsilon$-values. $V$, voltage variable, $w$, slow time scale variable, $I_{rheo}$, rheobase, $\epsilon$, time-scale coupling factor, $\tau_w$, trace, $\Delta$, determinant. }
	\label{fig:overview_models_TDdiagram}
\end{figure}

%mFHN
Recent studies of \cite{drion2012novel} and \cite{franci2012organizing} showed that the classical FHN model can be further generalised by incorporating a positive feedback (regenerative) in the slow conductance, $g_s$ \cite{drion2015dynamic}, in addition to the already known negative feedback (restorative). This discovery is a consequence of reducing the Hodgkin-Huxley model with calcium, which acts as a slow positive feedback necessary for bursting behaviour. The incorporation of the slow positive feedback in $g_s$ uncovered a lower branch of the $V$-nullcline. This further generalises the FHN model into the mirrored FHN (mFHN) model, where an additional co-dimension 3 pitchfork bifurcation divides the system into five excitability types \cite{franci2012organizing}. When the recovery variable, $w$, is only restorative on the membrane potential variations, the qualitative properties are similar to the classical FHN model. The transition between the regenerative and restorative property of slow gating ion channels underlies a transcritical bifurcation, which also occurs in many other conductance based models (CBM) and is therefore an essential property of the model \cite{franci2013balance}. In this work, we show that the incorporation of the regenerative feedback results in a dynamical behaviour that explains the observations during Type I and Type II excitability. The uncovering of the lower branch of the $V$-nullcline results in the existence of a funnel where the vector field, and thus $\epsilon_{crit}$, is able to maintain a SNIC bifurcation while the system is able to achieve a zero-frequency onset. In addition, the recovery variable $w$ is now able to capture multiple dynamical features of the system as the position of the $w$-nullcline affect the type and amount of open slow gating ion channels. This allows for the creation of a situation in the parameter space where it is possible to have a small transmembrane depolarising current during ISI, while the system undergoes SNIC bifurcation and the vector field allows for zero-frequency onset, which are all the requirements for Type I excitability. Using the concept of a positive and negative feedback balance in the conductance of slow gating ion channels, i.e. $g_s$, we found that these channels have a threshold state that translates to the mathematical value for which there is a bifurcation change, i.e. $\epsilon_{crit}$. Understanding how mathematical analysis compares to the neuronal physiology is crucial for the relevance of a model. This not only allows for more insight into neuronal dynamics and signal representation in firing rates, but also provides building blocks for understanding the role of neuronal excitability in memory formation. \par

%%%%%%%%%%%%%%%%%%%%%%%%%%%%%%%%%%%%%%%%%%%%%%%%%%%
%%%%%%%%%%%%%%%%% 			Results						%%%%%%%%%%%%%%%%%%%
%%%%%%%%%%%%%%%%%%%%%%%%%%%%%%%%%%%%%%%%%%%%%%%%%%%
\section{Results}
\subsection{The effect of time-scale separation on excitability type}

To model both Type I and Type II excitability, the system should have three fixed points: a left stable fixed point, a saddle and an unstable right fixed point. Although Type II excitability can be modelled in a system with one fixed point,
Type I excitability is only possible in a system that allows for infinite period cycle (i.e. zero-frequency onset), which can only occur with three fixed points, such as in the classical FitzHugh-Nagumo model (FHN) with a sigmoidal slow gating nullcline  (eqn. \ref{eq:FHN_eqn}) and the mirrored FitzHugh-Nagumo (mFHN)  (eqn. \ref{eq:mFHN_eqn}). \par

\begin{equation} \label{eq:FHN_eqn}
\begin{split}
&\textit{The classical FHN model:}\\
&\dot{V} = V - \frac{V^3}{3} - w + I_{app}\\
&\dot{w} = \epsilon (w_{\infty} (V-V_0) - w)\\
\end{split}
\end{equation}

\begin{equation} \label{eq:mFHN_eqn}
\begin{split}
&\textit{The mFHN model:}\\
&\dot{V} = V - \frac{V^3}{3} - w^2 + I_{app}\\
&\dot{w} = \epsilon (w_{\infty} (V-V_0) + w_0 - w)
\end{split}
\end{equation}

\noindent
In both FHN as mFHN, $w_{\infty} (V)$ is the standard Boltzman activation function:

\begin{equation} \label{Boltzmann_eqn}
w_{\infty} (V) := \frac{2}{1+e^{-5V}},
\end{equation}

\noindent
where $V$ is the fast voltage variable, and $w$ the recovery variable. In the mFHN model, the slow conductance is regenerative when $w_0 < 0$ and restorative when $w_0 > 0$. In both models, the evolution of the fast variable $V$ is much faster than the evolution of the slow variable, $w$. This is reflected in the time-scale coupling factor, $\epsilon = \frac{1}{\tau_w}$, where $\tau_w >> 1$ and $\epsilon << 1$. A large time-scale coupling factor means that $\tau_w$ has a large value, and thus $\epsilon$ has a small value.\par 
To study the different types of excitability, we can analyse the dynamical systems by plotting the dynamical variables $V$ and $w$ in the phase plane \cite{strogatz2018nonlinear}, \cite{izhikevich2007dynamical}, \cite{gerstner2002spiking} (see fig. \ref{fig:overview_models_TDdiagram}).
The important difference between Type I and Type II excitability is the process of spike initiation. Spike initiation occurs at the local minimum of the $V$-nullcline. Changing the applied current, $I_{app}$, either results in the change of stability of the left fixed point via an Andronov-Hopf bifurcation, or the collision of the left fixed point with the saddle via a saddle-node on invariant cycle (SNIC) bifurcation. Canonically, each bifurcation type explains a class of excitability: Type I is caused by SNIC; Type II is the result of Adronov-Hopf. Once the left fixed point is destabilised via Andronov-Hopf bifurcation, spiking ensues, and the collision with the saddle does not affect the spiking. However, as long as the left fixed point is kept stable it will collide with the saddle via a SNIC bifurcation. Therefore, the stability of the left fixed point decides whether the system changes from rest to spiking before or upon the collision with the saddle. \par 

% effect epsilon on stability left-fixed point
Whether the left fixed point remains stable is due to the vector field, which, on its turn, is affected by the time-scale coupling factor $\epsilon$ (fig. \ref{fig:overview_models_TDdiagram}). Therefore, changing  $\epsilon$ influences the stability of the left fixed point. The effect of $\epsilon$ on the left fixed point is explained by looking at the linearisation at the fixed point, i.e. the Jacobian, $J$ (eqn. \ref{eq:Jacobian}). The general requirement for a bifurcation is that $\Delta(J) = 0$. The difference between SNIC and Hopf bifurcation is that during Hopf bifurcation the eigenvalues are complex conjugates that cross the imaginary axis upon bifurcation (see \cite{strogatz2018nonlinear}, pg. 248). Therefore, the radical in the quadratic equation is zero upon bifurcation, and so $\tau(J) = 0$. Looking at the Jacobian, the $\epsilon$-values cancels out when calculating the determinant, however, the trace is still dependent on $\epsilon$. Therefore, changes in $\epsilon$ will only affect the trace of the system, i.e. the destabilisation of the left fixed point. Changes in $\epsilon$ correspond to the movement in the $\tau$-direction in the trace-determinant (TD) diagram (node $\rightarrow$ spiral $\rightarrow$ node, fig. \ref{fig:overview_models_TDdiagram}), which means keeping the left fixed point stable or change into a spiral.

\begin{equation}
J =
\begin{bmatrix}
\frac{\partial \dot{V}}{\partial V} & \frac{\partial \dot{V}}{\partial w}  \\
\frac{\partial \dot{w}}{\partial V} & \frac{\partial \dot{w}}{\partial w} \\
\end{bmatrix}
= 
\begin{bmatrix}
1-V^2 & -2w \\ 
\frac{\partial{w_{\infty}}}{\partial{V}} (V-V_0)  & - \epsilon \\
\end{bmatrix}
\label{eq:Jacobian}
\end{equation}

\begin{align}
& \tau(J) = (1-V^2) - \epsilon \\
& \Delta(J) = ((V^2-1) - (-2w \frac{\partial{w_{\infty}}}{\partial{V}} (V-V_0))
\end{align}

\begin{figure}[h]
	\begin{minipage}[c]{0.67\textwidth}
		\includegraphics[width=\textwidth]{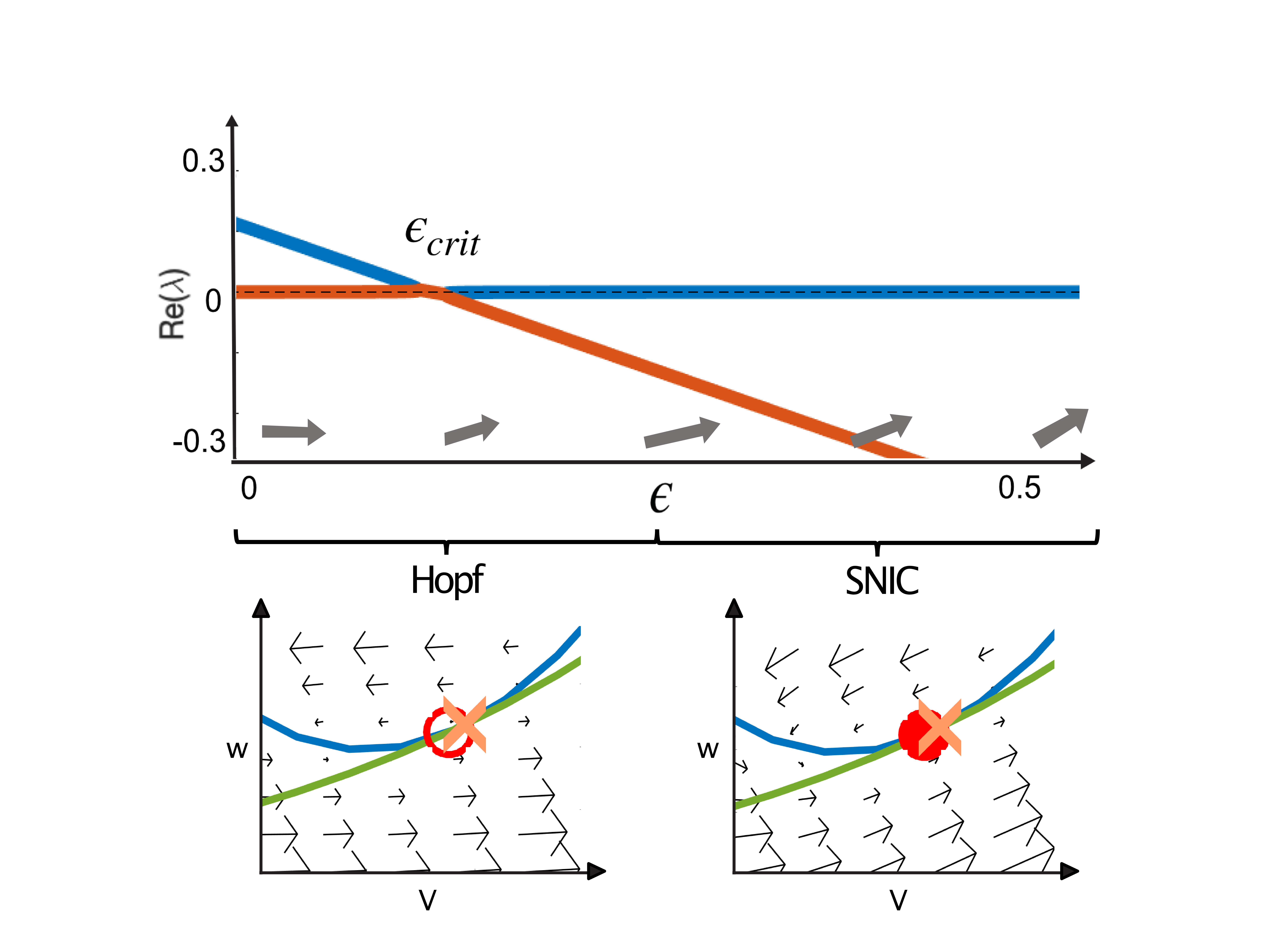}
	\end{minipage}\hfill
	\begin{minipage}[c]{0.3\textwidth}
		\caption{\textbf{Bifurcation change at $\epsilon_{crit}$}. The real part of the eigenvalues, $Re(\lambda)$, of the left fixed point are plotted for different $\epsilon$-values. Changing the $\epsilon$-value from small to large, results in an unstable left fixed point ($Re(\lambda) > 0$ for $\epsilon = \epsilon_{crit}$) to a stable left fixed point ($Re(\lambda) < 0$ for $\epsilon = \epsilon_{crit}$), respectively. Here, the left fixed point in the mFHN model is used, however, the left fixed point in the classical FHN model is similar. Parameters: $V_0 = -0.2$, $w_0 = 0.2$ and $I_{app} = I_{SN}-1e^{-14}$, where $I_{SN} = 0.7248$.} 
		\label{fig:bfnChange}
	\end{minipage}
\end{figure}

% effect of saddle on non-zero part w-nullcline
The effect of $\epsilon$ on the left fixed point just before collision with the saddle is shown in figure \ref{fig:bfnChange}. The figure shows a zoom of the region around the local minimum of the $V$-nullcline, which shows that the saddle is located on the non-zero slope part of the $w$-nullcline. This is due to the sigmoidal shape of the $w$-nullcline, and occurs in both the FHN (eq. \ref{eq:FHN_eqn}) and the mFHN model (eq. \ref{eq:mFHN_eqn}). The location of the saddle at the non-zero slope part is particularly important for Type I excitability, as the saddle location in SNIC is the threshold for firing onset, i.e. $V_{th}$. During Type I excitability, the continuously firing at a low rate is maintain by an extremely small transmembrane depolarising current during the inter-spike interval (ISI) \cite{khaliq2008dynamic}. The location of $V_{th}$, and therefore the saddle, should be located at non-zero slope part, so that some slow gating ion channels are open at $V_{th}$. Thus both a stable left fixed point upon collision with the saddle, and the location of the a saddle at the non-zero part of the $w$-nullcline is a requirement for Type I excitability. \par  

% effect of saddle on off-local minimum => e_crit
Due to the sigmoidal shape, the saddle is not only located on the non-zero slope part of the $w$-nullcline, but also moved to the right of the $V$-nullcline local minimum, instead of being at the local minimum of the $V$-nullcline (see fig. \ref{fig:bfnChange}). Therefore, there exist a regime between the local minimum and the saddle where $\epsilon$ allows for the vector field to be tangential to the left fixed point \textit{before} collision with the saddle. In this regime, the left fixed point is either destabilised to an unstable node ($Re(\lambda) > 0$), or, as long as the vector field is oblique enough, the left fixed point remains its stability ($Re(\lambda) < 0$). This change of stability is continuous and therefore there is an $\epsilon$-value such that $Re(\lambda) = 0$. This is the $\epsilon_{crit}$-value, where Hopf = SNIC, and thus the value for which a change of bifurcation type occurs, i.e. a change from Hopf bifurcation to SNIC bifurcation.\par

% effect of epsilon on minimum spiking rate
As shown in figure \ref{fig:overview_models_TDdiagram}, the global effect of $\epsilon$ in the vector field is its effect on the spiking rate. A small $\epsilon$-value, and therefore a large time-scale separation $\tau_w$, results in a long ISI and thus a low spiking frequency. In order to obtain a SNIC bifurcation that typifies Type I excitability, the system should allow for the property of zero-frequency onset when the system undergoes SNIC bifurcation. In other words, the $\epsilon_{crit}$-value needs to create a situation in which the left fixed point remains stable, while the vector field results in a large ISI, and thus the time-scale separation, $\tau_w$, is infinite. This situation is when $\epsilon_{crit}$ is minimal, which we can study using singular limit analysis, i.e. when $\epsilon = 0$ \cite{jones1995geometric}. In summary, the aim is to find a system that allows for both Type I and Type II excitability, in which there must be a regime that allows for stability of the left fixed point and zero-frequency onset. \par

\begin{figure}[h!]
	\begin{minipage}[c]{0.67\textwidth}
		\centering
		\includegraphics[width=0.6\textwidth]{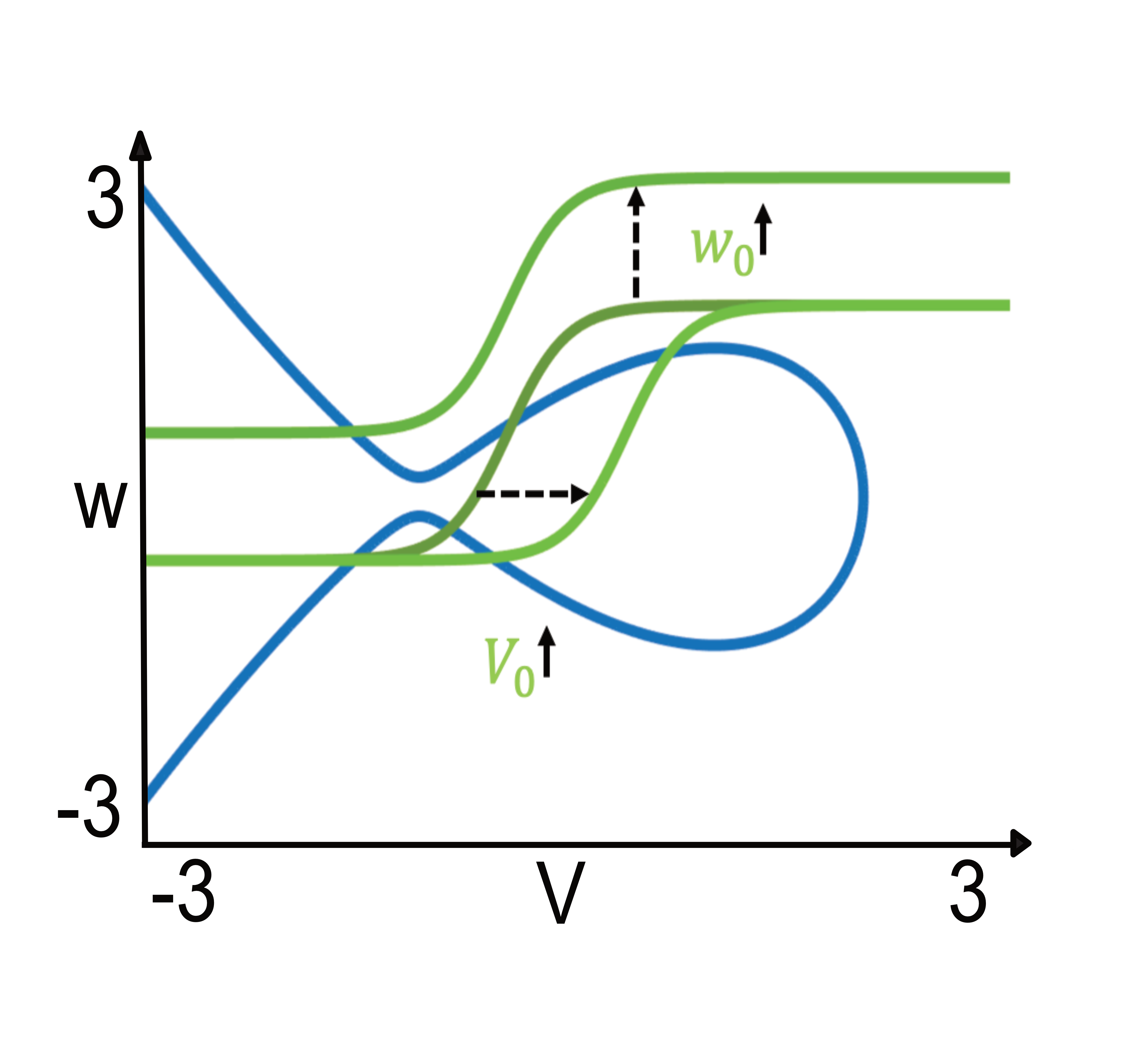}
	\end{minipage}\hfill
	\begin{minipage}[c]{0.3\textwidth}
		\caption{\textbf{Effect of parameters $V_0$ and $w_0$ in the phase plane.} Increasing $V_0$ moves the $w$-nullcline to the right, which moves the saddle to the more horizontal part of the $w$-nullcline. Changing $w_0$ moves the $w$-nullcline up or down and allows for the slow gating ion channels to be either restorative (upper branch of $V$-nullcline) or regenerative (lower branch of $V$-nullcline)}.
		\label{fig:effect_parameters}
	\end{minipage}
\end{figure}

\begin{figure}[h]
	\centering
	\includegraphics[width=0.9\linewidth]{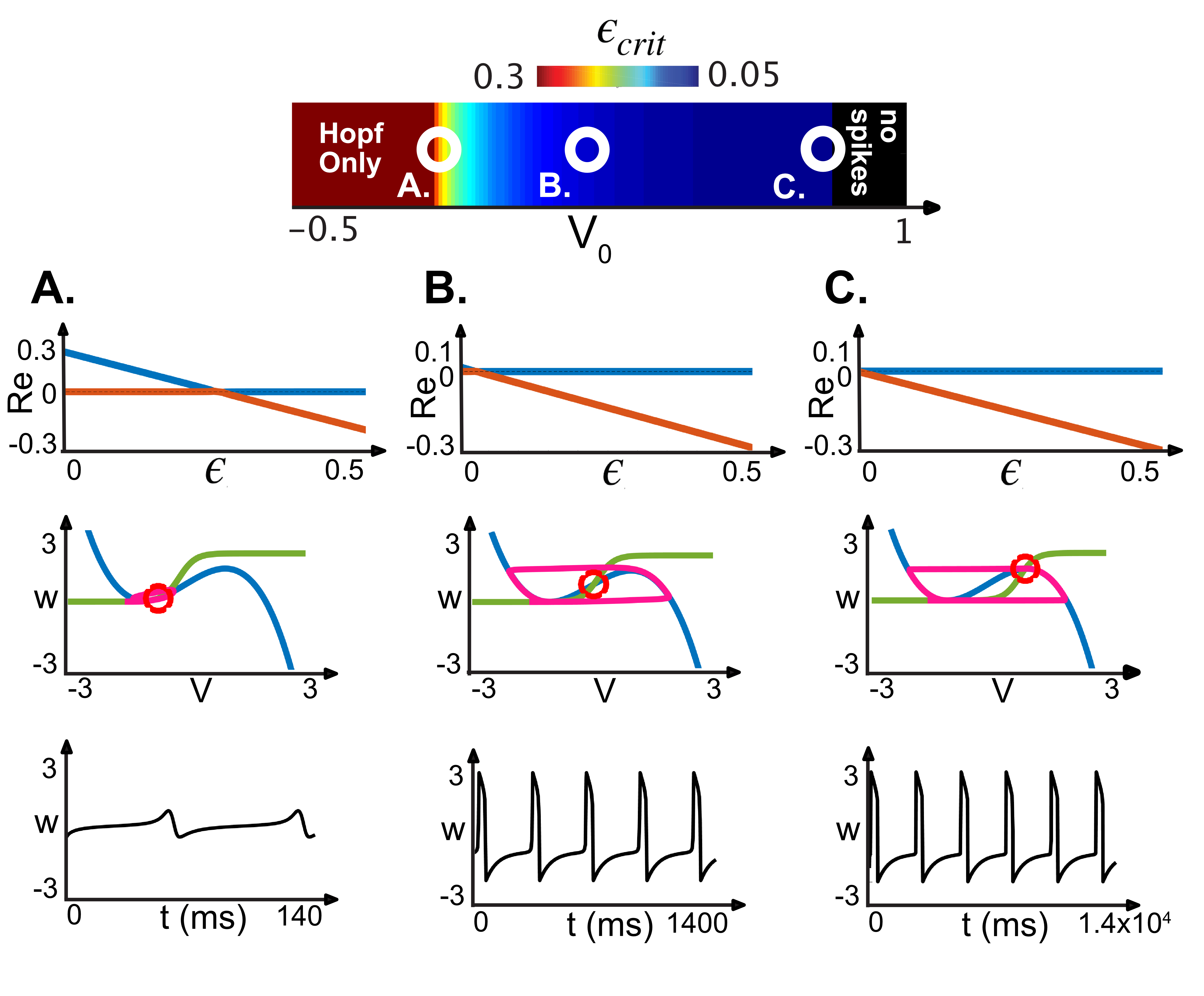}
	\caption{\textbf{The effect of parameter $V_0$ on $\epsilon_{crit}$ in the classical FHN model}. Changing $V_0$ moves the $w$-nullcline and affect the $\epsilon_{crit}$-value. In order to obtain Type I excitability, $\epsilon_{crit}$ should be as low as possible (dark blue). When the $\epsilon_{crit}$ is relatively high, the spike height decreases (see \textbf{A}). In \textbf{B}, the $\epsilon_{crit}$-value is not at the lowest value and therefore SNIC bifurcation doesn't lead to a frequency that has an infinite period. For the lowest $\epsilon_{crit}$-value, as in \textbf{C}, the saddle point is at the horizontal part of the $w$-nullcline, causing the slow gating ion channels to be at steady state. Parameters: (\textbf{A}) $V_0$ = -0.14; $\epsilon_{crit}$ = 0.326;  (\textbf{B}) $V_0$ = 0.18; $\epsilon_{crit}$ = 0.03; (\textbf{C}) $V_0$ = 0.82; $\epsilon_{crit}$ = 0.001.}
	\label{fig:FHNnc_epsilon_crit}
\end{figure}

\subsection{Incorporating a slow regenerative parameter in phenomenological models allows for Type I excitability}
\begin{figure}[h]
	\begin{minipage}[c]{0.67\textwidth}
		\includegraphics[width=\textwidth]{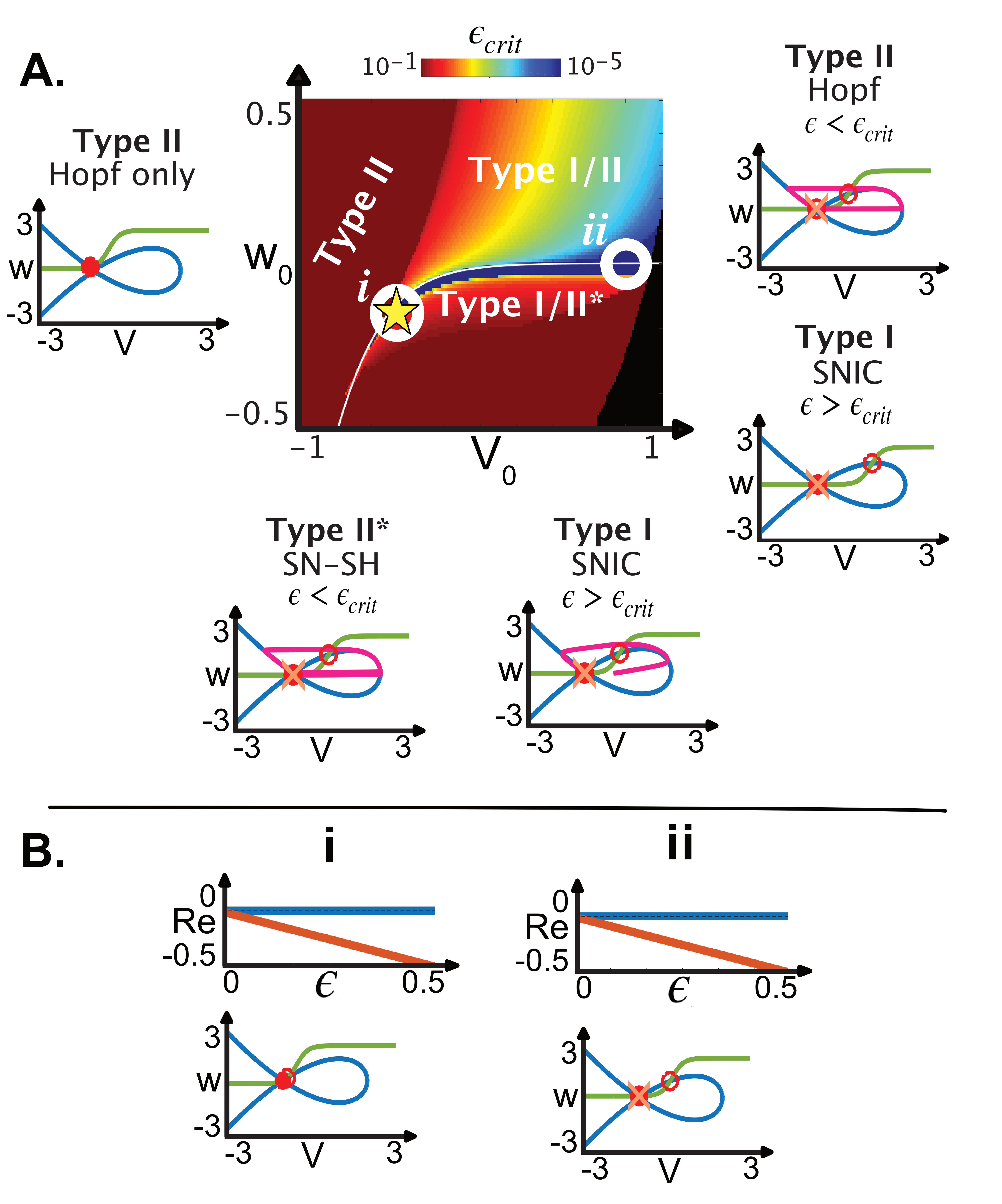}
	\end{minipage}\hfill
	\begin{minipage}[c]{0.3\textwidth}
		\caption{\textbf{The effect of parameters $V_0$ and $w_0$ on $\epsilon_{crit}$ in the  mFHN model}. In \textbf{A.}, the $V_0, w_0$-space is separated in different excitability types via a pitchfork bifurcation (star), where the transcritical bifurcation (white line) separates the space into positive and negative feedback in the slow gating ion channels. The rainbow colour scale indicates the $\epsilon_{crit}$-value. Small $\epsilon_{crit}$-values, $\epsilon_{crit} =10^{-5}$, occur around the transcritical line (indicated in dark blue). The phase portraits of each excitability type is depicted around the parameter plot. The excitability types in the parameter space are: Type II (Hopf only) with one fixed point in the system that is destabilised via subcritical Hopf bifurcationn (dark red area); the region Type I/II (rainbow colours, above the transcritical bifurcation line), where the system has three fixed points and the left fixed point is either destabilised via SNIC ($\epsilon > \epsilon_{crit}$) or subcritical Hopf ($\epsilon < \epsilon_{crit}$); and the region Type I/II* (rainbow colours, below the transcritical bifurcation line), where the excitability class either underlies a saddle-node on saddle-homoclinic bifurcation (SN-SH, $\epsilon < \epsilon_{crit}$) or a SNIC, which also depends on the $\epsilon$-value.  \textbf{B.} indicates the situation where SNIC occurs in the singular limit. At (\textbf{i}), the parameters are close to the pitchfork bifurcation. Here, it is possible to locate the SNIC bifurcation at the non-zero part of the $w$-nullcline. The coordinates in (\textbf{ii}) resembles the situation in the classical FHN model, therefore, no Type I excitability in the singular limit is possible as the saddle point is at the horizontal part of the $w$-nullcline.}
		\label{fig:mFHN_epsilon_crit}
	\end{minipage}
\end{figure}

In order to obtain Type I excitability, there should be a configuration in the singular limit where the $\epsilon_{crit}$ allows for a stable left fixed point and an infinite time-scale separation. The $\epsilon_{crit}$-value is influenced by changes in the parameter $V_0$, which is the half-value activation voltage of the slow ion channels (fig. \ref{fig:FHNnc_epsilon_crit}, \ref{fig:mFHN_epsilon_crit}). $V_0$ is a parameter in both the FHN and the mFHN model. In the phase portrait, changing $V_0$ moves the $w$-nullcline in the horizontal direction, which changes the position of the saddle, i.e. $V_{th}$, on the $w$-nullcline, and thus affects the amount of open slow gating ion channels at the onset of firing. The effect of changing $V_0$ on $\epsilon_{crit}$ in the FHN model is shown in figure \ref{fig:FHNnc_epsilon_crit}. This shows that for a relatively high $\epsilon_{crit}$-value (bright red) the spike height decreases, whereas the lowest $\epsilon_{crit}$-value (dark blue) shows that the saddle point is at the horizontal part of the $w$-nullcline, resulting in the slow gating ion channels being at steady state. Mathematically, this shows that in the classical FHN model there is no configuration of the phase portrait that can lead to a SNIC bifurcation in the singular limit.\par
The difference between FHN and mFHN, is that mFHN incorporates an additional parameter $w_0$ in the recovery variable of the model. In the phase plane, changing $w_0$ results in the vertical movement of the $w$-nullcline, so that the collision with the saddle either occurs at the upper branch of the $V$-nullcline, reflecting negative feedback in the slow timescale, or at the lower branch of the $V$-nullcline, representing positive feedback \cite{franci2013balance}. The essential property of the mFHN model is the transcritical bifurcation, which governs the change of location of the saddle on either the upper branch or the lower branch. Therefore, the transcritical bifurcation is the switch between either negative feedback or positive feedback in the slow gating ion channels.  \par

% w_0 parameter in mFHN
Instead of one parameter, $V_0$, that influence the $\epsilon_{crit}$-value, the mFHN has the additional $w_0$ parameter that affects the $\epsilon_{crit}$-value by moving the $w$-nullcline up and down (fig.  \ref{fig:mFHN_epsilon_crit}). In mFHN, $V_0$ and $w_0$ can be chosen such that the system is close to the transcritical bifurcation (white line in fig. \ref{fig:mFHN_epsilon_crit}). At the transcritical bifurcation, the $V$-nullcline intersects itself, where the narrowing of the $V$-nullcline branches create a funnel. In this funnel, the $V$-components of the vectors are very small but non-zero, and thus the $\epsilon_{crit}$-values are low (dark blue in fig. \ref{fig:mFHN_epsilon_crit}). Locating the saddle close to the transcritical bifurcation forces the trajectory to go through this funnel upon SNIC bifurcation, and therefore allows for the period of spiking frequency to go to infinity, and thus meet the requirements for Type I excitability. \par  

% locations where Type I is in the singular limit
Another requirement for Type I excitability is the location of the saddle on the non-zero slope of the $w$-nullcline upon SNIC bifurcation. Therefore, looking at the parameter plot for which $\epsilon_{crit}$ is low shows two different situations (fig. \ref{fig:mFHN_epsilon_crit}\textbf{A}$i$ and $ii$) that can be studied using their phase plane (fig. \ref{fig:mFHN_epsilon_crit}\textbf{B}). The phase portrait of location $ii$, shows that the SNIC bifurcation, and therefore the location of the saddle, is located at the horizontal part of the $w$-nullcline. This is an identical situation to the FHN model, from which we know that in this situation the slow gating ion channels are at steady state, and thus does not allow for Type I excitability. On the other hand, at location $i$ SNIC bifurcation occurs at the non-zero part of the $w$-nullcline, and thus meets the physiological requirement for Type I excitability. The combination of infinite spiking frequency (low $\epsilon_{crit}$-value) and open slow gating ion channels is a configuration that persists in the singular limit. \par 

% the importance of the pitchfork bifurcation 
The parameter plot in figure \ref{fig:mFHN_epsilon_crit}\textbf{A} shows that location $i$ is around the pitchfork bifurcation. The pitchfork bifurcation in the mFHN is highly degenerate (codimension 3) and organises the parameter space into five different types of excitability: I, II, III, IV or II$*$, and V, see \cite{franci2012organizing}. Type III and Type V are less excitable variants of Type II and Type IV (II*), respectively. Type IV underlies a saddle-node on saddle-homoclinic (SN-SH) bifurcation and is characterised by bistability (see phase plane plots in \ref{fig:mFHN_epsilon_crit}\textbf{A} and supplementary material). The incorporation of a slow regenerative parameter, $w_0$, such as in the mFHN, does not only create a situation that reflects Type I excitability, but also indicates the type of bifurcation that underlies Type I excitability.

\subsection{The slow conducting ion channels at spiking threshold indicates the change in excitability type}
In biological neurons, many ion channels contribute to the different timescales to the excitability of the neuron. This general concept underlies the computation of the dynamic input conductances (DIC), which calculates the contribution of each ion channel to one of the three timescale, $g_f$, $g_s$ and $g_u$ \cite{drion2015dynamic}. Here we focus on $g_s$, as $g_f$ is responsible for the fast spike upstroke and therefore too fast to contribute to dynamical behaviour of excitability, and $g_u$ is mainly important for bursting behaviour. Drion (2015)
showed that $g_s$ drives the excitability change Type II – Type I – Type II*, where Type I excitability occurs for $g_s(V_{th}) \approx 0$, i.e the slow time-scale at threshold value is close to zero. In general, $g_s$ is the derivative of the slow current with respect to membrane potential (\ref{eq:gs}). In the phenomenological models, this is easily computed as the slow time-scale dynamics depend on a single variable, $w$.

\begin{equation}
g_s = -\frac{\partial{\dot{V}}}{\partial{w}} \frac{\partial{w_{\infty}}}{\partial{V}}
\label{eq:gs}
\end{equation} 

\begin{figure}[h]
	\begin{minipage}[c]{0.67\textwidth}
		\includegraphics[width=\textwidth]{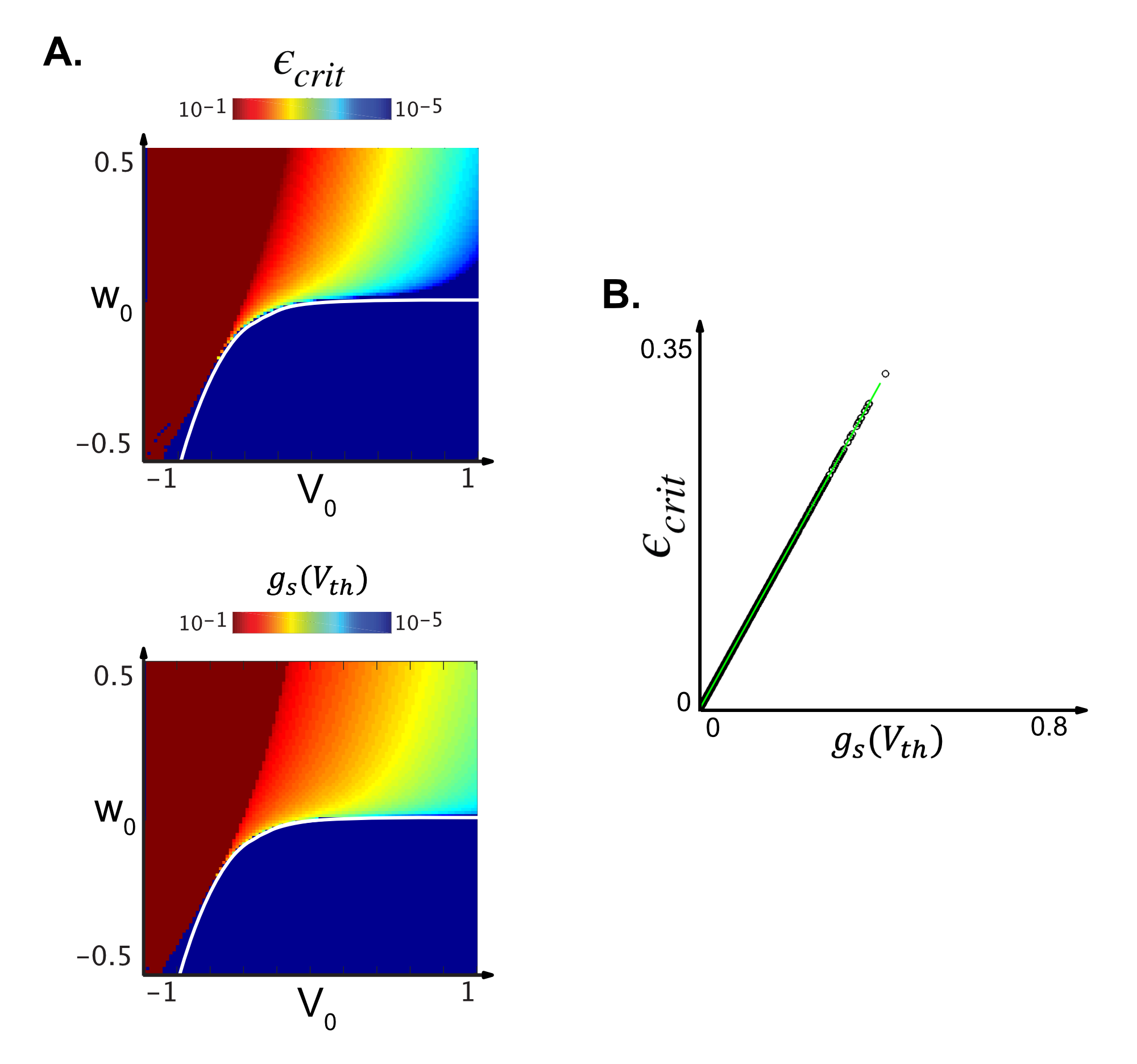}
	\end{minipage}\hfill
	\begin{minipage}[c]{0.3\textwidth}
		\caption{\textbf{Linear relationship between $\epsilon_{crit}$ and $g_s(V_{th})$}. \textbf{A.} shows that $g_s(V_{th}$ has a similar gradual change as $\epsilon_{crit}$. This similarity is also indicated by the linear relationship between the $\epsilon_{crit}$ and $g_s(V_{th})$-values (see  \textbf{B.}). Colour scales are in log$_{10}$-scale.}
		\label{fig:gsecrit}
	\end{minipage}
\end{figure}

\noindent
The terms on the right hand side implies that either the slope of the $w$-nullcline is almost zero at $V_{th}$ ( $\frac{\partial w_{\infty}}{\partial V} \rvert_{V_{th}} \approx 0$) or the derivative of the membrane current with respect to the slow gating variable is around zero at $V_{th}$ ($\frac{\partial \dot{V}}{\partial w} \rvert_{V_{th}} \approx 0$). We already know from the results of mFHN that when the saddle node is on the horizontal part of the sigmoidal $w$-nullcline, the ion-channels in the slow time-scale are at steady state. Therefore, $\frac{\partial w_{\infty}}{\partial V} \rvert_{V_{th}} \neq 0$. In the phenomenological models, a threshold is only measurable when the left fixed point remains stable, i.e. when SNIC bifurcation occurs. Therefore, $g_s(V_{th})$ is measurable as soon as the change in bifurcation occurs from Hopf to SNIC, which is at $\epsilon_{crit}$. Calculating the exact value of $g_s(V_{th})$-value in the mFHN model shows that $g_s(V_{th})$ relates linearly to $\epsilon_{crit}$ (see fig. \ref{fig:gsecrit}).\par

To prove this linear relationship, consider the mFHN planar model, eqn. \ref{eq:mFHN_eqn}, and the condition for which $\epsilon_{crit}$ occurs, i.e. SNIC=Hopf. The conditions for SNIC bifurcation is the general condition for bifurcation, i.e. $\Delta(J) = 0$, where $\Delta$ is the determinant and $J$ is the Jacobian. The Jacobian is given by

\[
J =
\begin{bmatrix}
\frac{\partial \dot{V}}{\partial V} & \frac{\partial \dot{V}}{\partial w}  \\
\frac{\partial{w}}{\partial V} & \frac{\partial \dot{w}}{\partial w} \\
\end{bmatrix}
= 
\begin{bmatrix}
1-V^2 & -2w \\ 
\frac{\partial{w_{\infty}}}{\partial{V}} (V-V_0) & - 1 \\
\end{bmatrix},
\]

\noindent
and thus

\begin{align*}
\Delta(J) &= \frac{\partial \dot{V}}{\partial V}  \frac{\partial \dot{w}}{\partial w} - \frac{\partial \dot{V}}{\partial w} \frac{\partial \dot{w}_{\infty}}{\partial V} = 0
\label{SN_det}
\end{align*}

\noindent
gives

\begin{equation}
\frac{\partial \dot{V}}{\partial V}  \frac{\partial \dot{w}}{\partial w} = \frac{\partial \dot{V}}{\partial w} \frac{\partial \dot{w}_{\infty}}{\partial V}.
\end{equation}

\noindent
The right hand side, $\frac{\partial \dot{V}}{\partial w} \frac{\partial \dot{w_{\infty}}}{\partial V}$, is equal to the definition $-g_s$ in eqn. \ref{eq:gs}. 

\begin{equation}
\frac{\partial \dot{V}}{\partial V}  \frac{\partial \dot{w}}{\partial w} = -g_s,
\end{equation}

\noindent
so that 

\begin{align*}
&(1-V^2)*(-1) = -g_s,\\
&V^2 - 1 = -g_s,\\
&g_s = 1 - V^2 . \numberthis
\label{gs_proof}
\end{align*}

\noindent
As described in \cite{strogatz2018nonlinear}, the conditions for Hopf bifurcation are $\tau(J) = 0$. This gives

\begin{align*}
&\tau(J) = \frac{\partial \dot{V}}{\partial V} + \frac{\partial \dot{w}}{\partial w} = 0,\\
& 1-V^2 - \epsilon_{crit} = 0,\\
&\epsilon_{crit} = 1-V^2.  \numberthis
\label{tepscrit_proof}
\end{align*}

\noindent
Taking the conditions for SNIC, eqn. \ref{gs_proof}, and Hopf, eqn. \ref{tepscrit_proof}, together this implies that

\begin{equation}
\epsilon_{crit} = 1-V^2 = g_s.
\end{equation}

\noindent
The proof shows that the value at $\epsilon_{crit} =  g_s(V_{th})$, thus the slow ion conductance at $V_{th}$ is the exact and an experimentally measurable value for which there is a bifurcation change. This is an important finding, as it shows that the slow conducting ion channels are responsible for the excitability change in neurons, and also relates the mathematical finding to an experimental measurable value of the slow conducting ion channels. \par

\subsection{How reduced conductance based models indicate the type of slow gating ion channel that is reponsible for the excitability change}

\begin{figure}[h]
	\centering
	\begin{subfigure}{.5\linewidth}
		\centering
		\includegraphics[width=0.8\linewidth]{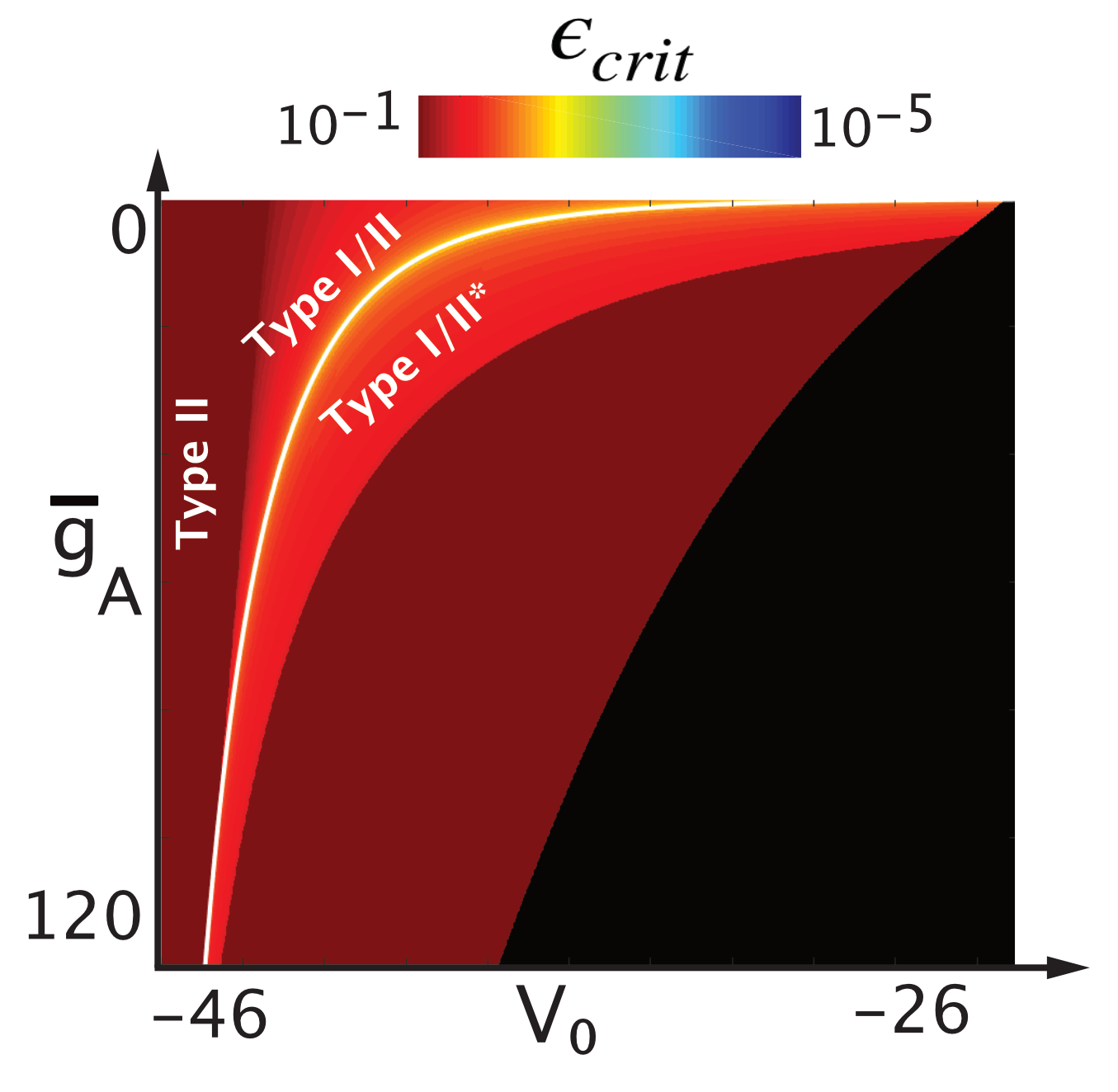}
		\caption*{\textbf{Reduced Connor-Stevens}}
		\label{fig:CSred_bifn_map}
	\end{subfigure}%
	\begin{subfigure}{.5\linewidth}
		\centering
		\includegraphics[width=0.8\linewidth]{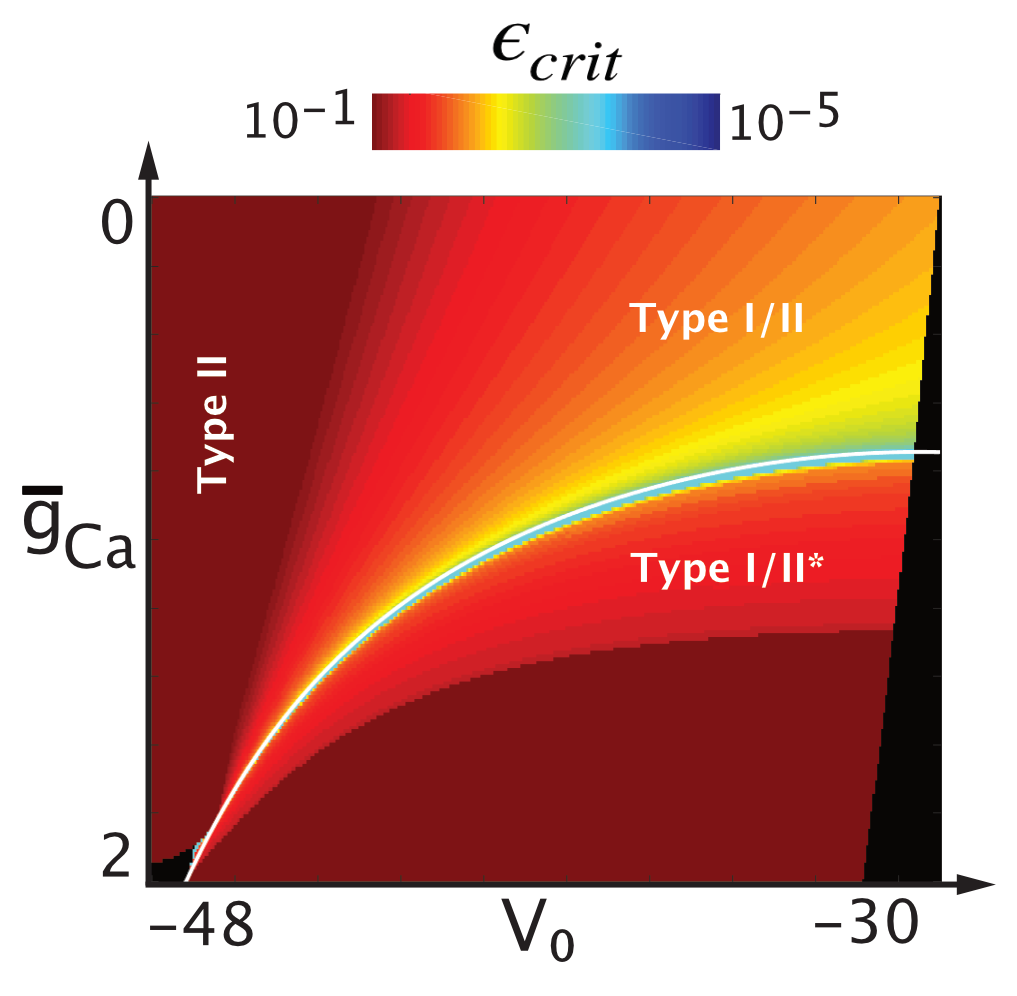}
		\caption*{\textbf{Reduced Hodgkin-Huxley with Ca$^+$}}
		\label{fig:HHCa_bifn_map}
	\end{subfigure}
	\caption{\textbf{$\epsilon_{crit}$ in the parameter space of reduced conductance based models.} The left is parameter plot ($\bar{g}_A, V_0$) is the reduced Connor-Stevens (CS) model; the right parameter plot ($\bar{g}_{Ca}, V_0$) is the reduced Hodgkin-Huxley with Ca$^+$. In both plots, the vertical axis ($\bar{g}_A, \bar{g}_{Ca}$ respectively) is the slow gating variable that corresponds to $w_0$ in the mFHN model. Transcritical bifurcation (white line); Type II only (dark red), Type I/II (rainbow colours, above transcritical bifurcation line), and Type I/II* (rainbow colours, below transcritical bifurcation line), right fixed point stable (black region).}
	\label{fig:CBMred_bifn_map}
\end{figure}

\noindent
The importance of mathematical findings in phenomenological models is that these are generalisable and translate to the more physiological reflecting conductance based models (CBMs). CBMs describe the dynamical interaction between the membrane potential and gating variables with parameters that are recorded using experiments. Here, we compare the findings from the mFHN model to the reduced CBMs called the reduced Connor-Stevens (CS) \cite{connor1971prediction}, \cite{connor1977neural}, \cite{drion2015ion} and the Hodgkin-Huxley with added calcium (HH-Ca$^+$) \cite{hodgkin1952quantitative}, \cite{drion2012novel}. \par
In the mFHN, we proved the relationship $\epsilon_{crit} =  g_s(V_{th})$. This relationship is also true for the reduced CBMs, and from this relationship we can derive an explicit notion of the time-scale coupling factor $\epsilon$ for each reduced CBM. Here, we consider the general system for reduced CBMs,

\begin{align}
&\dot{V} = \frac{1}{C} (\bar{g}_x m_x^a h_x^b (V-E_x) + I_{app}),\\
&\dot{n} = \frac{1}{\tau_n} n_\infty - n.
\end{align}

\noindent
The time dependencies can be taken outside the partial derivatives, and so we can rewrite the Jacobian as

\[
J =
\begin{bmatrix}
\frac{1}{C} \frac{\partial \dot{V}}{\partial V} & \frac{1}{C} \frac{\partial \dot{V}}{\partial n}  \\
\frac{1}{\tau_n} \frac{\partial n_\infty}{\partial V} & \frac{1}{\tau_n} \frac{\partial \dot{n}}{\partial n} \\
\end{bmatrix}
\].

\noindent
In the condition for SNIC, $det(J) = 0$, the time dependency and capacitance vanish.

\begin{align*}
&(\frac{1}{C} \frac{\partial \dot{V}}{\partial V}) (\frac{1}{\tau_n} \frac{\partial \dot{n}}{\partial n}) = (\frac{1}{C} \frac{\partial \dot{V}}{\partial n}) (\frac{1}{\tau_n} \frac{\partial n_\infty}{\partial V}), \\
&(\frac{\partial \dot{V}}{\partial V}) (\frac{\partial \dot{n}}{\partial n}) = (\frac{\partial \dot{V}}{\partial n}) (\frac{\partial n_\infty}{\partial V}). 
\end{align*}

\noindent
Using the expression of $g_s$ (eq \ref{eq:gs}) we obtain

\begin{align*}
&(\frac{\partial \dot{V}}{\partial V}) (\frac{\partial \dot{n}}{\partial n}) = -g_s, \\
&(\frac{\partial \dot{V}}{\partial V}) (-1) = -g_s, \\
&\frac{\partial \dot{V}}{\partial V} = g_s.  \numberthis
\label{eq:CBM_SNIC_gs}
\end{align*}

\noindent
In the conditions for Hopf bifurcation, $\tau = 0$, the time dependency and conductance expression remain

\begin{figure}[h]
	\includegraphics[width=\linewidth]{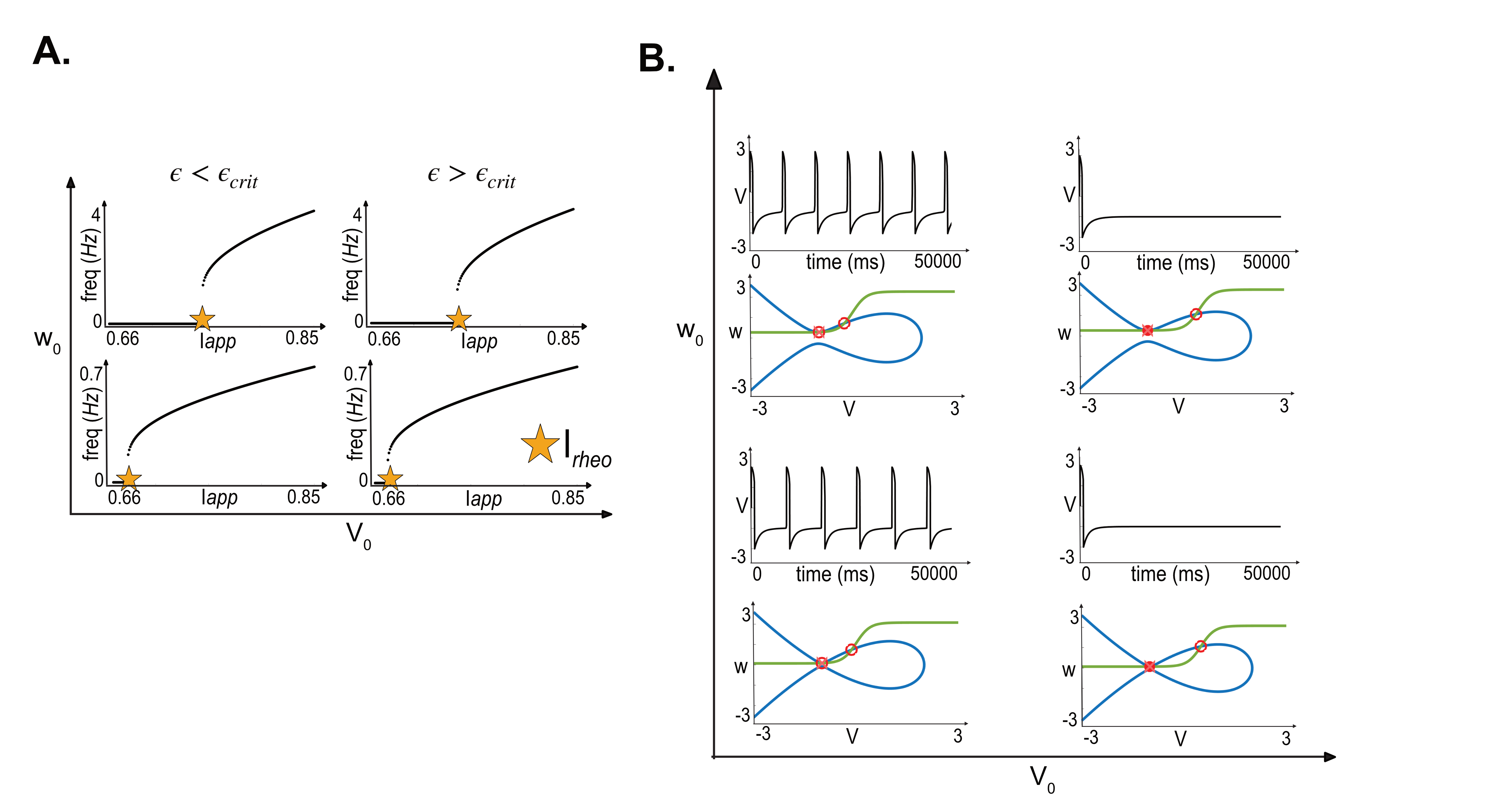} 
	\caption{\textbf{Local effect of parameters $V_0$ and $w_0$ on the frequency-current curve.} \textbf{A} indicates the local effect on the FI curve upon changes in $V_0$ and $w_0$. The $\epsilon$-value is chosen such that in the left quadrant, the system will undergo a Hopf bifurcation, while in the right quadrant the system will undergo a SNIC bifurcation. \textbf{B} shows the simulations that correspond to the same parameters, $V_0, w_0$, as in \textbf{A}. The $\epsilon$-values are chosen not far from the $\epsilon_{crit}$-value.}
	\label{fig:FHNm_effect_V0_w0_local}
\end{figure}

\begin{align*}
&\frac{1}{C} \frac{\partial \dot{V}}{\partial V} + \frac{1}{\tau_n} \frac{\partial \dot{n}}{\partial n} = 0,\\
&\frac{1}{C} \frac{\partial \dot{V}}{\partial V} + \frac{1}{\tau_n} (-1) = 0,\\
&\frac{1}{C} \frac{\partial \dot{V}}{\partial V} = \frac{1}{\tau_n}, \\
&\frac{\partial \dot{V}}{\partial V} = \frac{C}{\tau_n}.  \numberthis
\label{eq:CBM_Hopf_gs}
\end{align*}

\noindent
As before, to obtain the critical value we need to equate conditions eqn. \ref{eq:CBM_SNIC_gs} and eqn. \ref{eq:CBM_Hopf_gs}, i.e. Hopf = SNIC,

\begin{equation}
g_s = \frac{\partial \dot{V}}{\partial V} = \frac{C}{\tau_n}.
\end{equation} 

\noindent
And as the change of excitability type occurs at threshold value, which is the saddle point, the equation becomes

\begin{equation}
g_s(V_{th}) = \frac{C}{\tau_n}\Bigr\rvert_{V_{th}}  = \epsilon_{crit}, 
\end{equation}

\noindent
where $\frac{C}{\tau_n}\Bigr\rvert_{V_{th}}$ is the size of the slow current and corresponds to the critical timescale factor $\epsilon_{crit}$ in the mFHN model. \par
The calculation of $\epsilon_{crit}$ in both the reduced CS and HH+Ca$^+$ model is shown in figure \ref{fig:CBMred_bifn_map}. The slow conductances of both reduced CS and HH+Ca$^+$, $g_A$ and $g_{Ca}$ respectively, corresponds to $w_0$ from the mFHN model. This means that the reduced CBMs indicates the type of slow gating ion channel that is responsible for the change in excitability. Similar to the mFHN model, SNIC bifurcation can only occur in the singular limit around the transcritical bifurcation line (white line). This divides the parameter space into the same excitabilities classes as in mFHN, i.e. Type II only, Type I/II (above transcritical bifurcation line), and Type I/II* (below transcritical bifurcation line). This shows that the findings in the mFHN model are translatable to CBMs, and thus the contribution of different types of slow gating ion channels can be extracted and modelled in order to look for their effects in neuronal excitability.

\begin{figure}[h]
	\centering
	\includegraphics[width=1.0\linewidth]{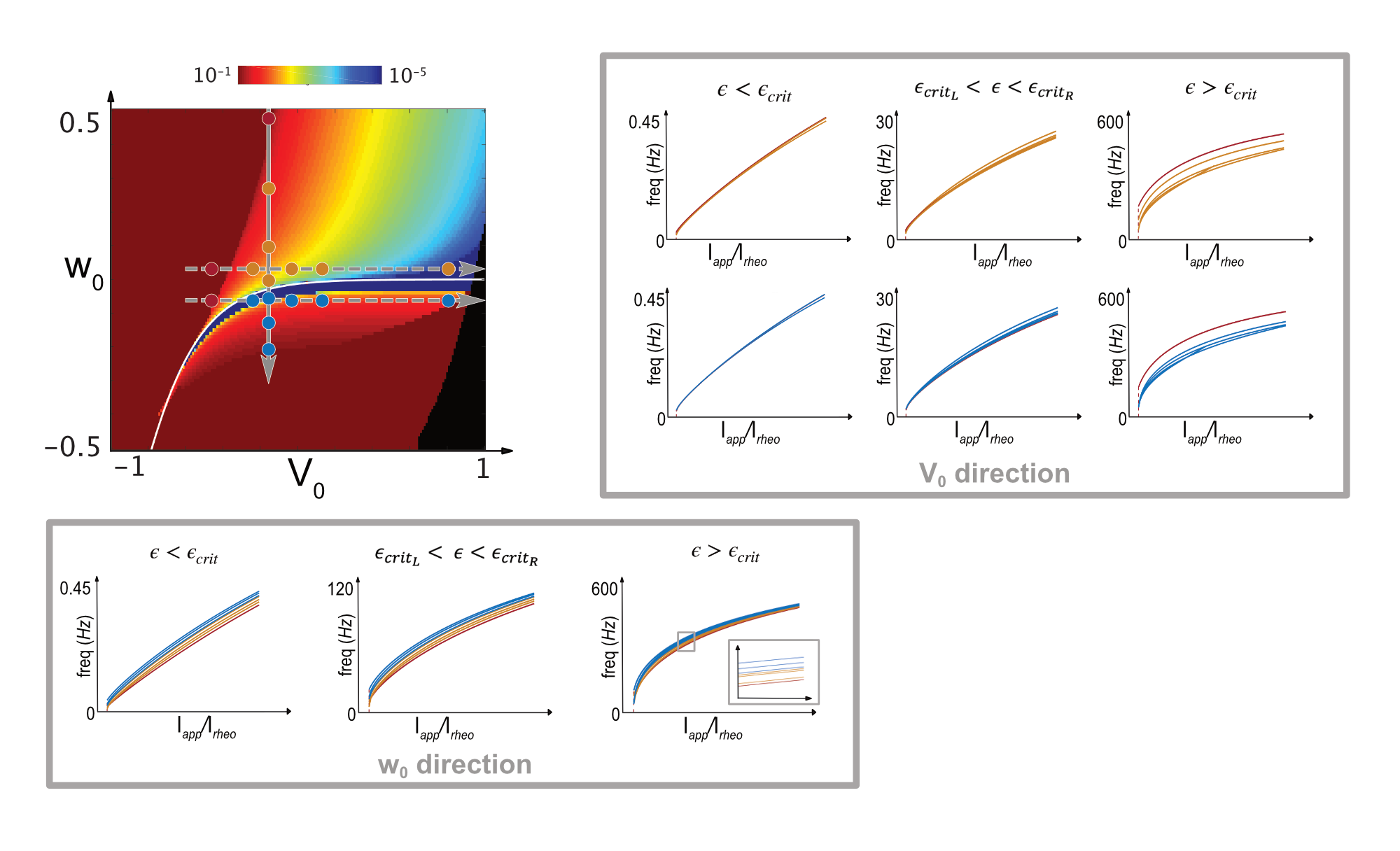}
	\caption{\textbf{Global effect of parameters $V_0$ and $w_0$ on the frequency-current curve.} On the left side, the parameter space indicates the location of the coordinates chosen on the $w_0$-path (vertical arrow) and the $V_0$-path (two horizontal arrows). A path in a parameter direction indicates that one parameter is kept stable, while the other parameter varies. Each location is simulated for Hopf bifurcation ($\epsilon < \epsilon_{crit}$), when halve of the coordinates undergoes Hopf and halve of the coordinates changes state via SNIC ($\epsilon_{{crit}_L} < \epsilon < \epsilon_{{crit}_R}$), and SNIC bifurcation  ($\epsilon > \epsilon_{crit}$). The overall effect of changing the $\epsilon$-value shows that the frequency increases (see fig. \ref{fig:overview_models_TDdiagram}). Decreasing $w_0$ (lower panel) increases frequency, while increasing $V_0$ keeps the frequency of the FIs similar. Red dots: system with one fixed point; orange dots: parameter values in system with three fixed points and above transcritical bifurcation line; blue dots: parameter values in system with three fixed points and below transcritical bifurcation line. Statistics of FI curves are in Supplementary Material.}
	\label{fig:FHNm_effect_V0_w0_global}
\end{figure}

\subsection{The local and global parameters that influence the FI curve signature for each excitability class.}

% what we know so far: Type I only possible when e_crit is low. e_crit is measurable at gs(Vth), but how to know when you're at the PF region when using FI.
In the previous sections, we showed that the canonical allocation of excitability class solely to bifurcation type is incomplete. This is also shown in figure \ref{fig:FHNm_effect_V0_w0_local}, where the change in underlying bifurcation does not result in a measurable difference in the FI curve. To meet the requirements for Type I bifurcation, we need a low $\epsilon_{crit}$-value, i.e. a large time-scale coupling, and a small transmembrane depolarizing current during ISI. In the mFHN, this means that we need to choose $V_0$ and $w_0$ such that we are close to the pitchfork bifurcation. The aim is to translate these conditions to experimentally measurable values. We already showed that $\epsilon_{crit} = g_s(V_{th})$, thus experimentally we can measure and tune the $g_s(V_{th})$ in order to get a low value. However, how can we translate the $V_0$, $w_0$ coordinates at pitchfork bifurcation to study Type I excitability using experimentally obtained FI curves? \par

% Local and global effects parameters $V_0$ and $w_0$
Figure \ref{fig:FHNm_effect_V0_w0_local}\textbf{A} shows the local effect of the $V_0$ and $w_0$ parameters on the FI curve. Each path contains similar values for the parameters, although the location in the parameter space is different. Locally, $w_0$ causes the shift of $I_{rheo}$ to the right. The rheobase, $I_{rheo}$, is the minimal current amplitude needed to depolarise the post-synaptic membrane to elicit an action potential. Changes in $V_0$ hardly moves the $I_{rheo}$ and thus results in similar looking FI curves. Besides these local effects, the FI curve is globally characterised by the overall change in frequency. In figure \ref{fig:FHNm_effect_V0_w0_global}, the FI curves are normalised for the $I_{rheo}$-value and show the global effect of $V_0$ and $w_0$ for different $\epsilon$-values. In general, an increase in $\epsilon$ results in an increase in frequency (see fig. \ref{fig:overview_models_TDdiagram}). The additional global effect of $w_0$ shows that the down-wards vertical movement for similar $w_0$-values increases the frequency, which is maintained for all $\epsilon$-values (fig. \ref{fig:FHNm_effect_V0_w0_global}, lower box). This effect is present for both a system with one fixed point (red curve/dot) and a system with three fixed points (blue and orange curves). Moving in the $w_0$-direction also shows that the difference between the FI curves are relatively stable for all $\epsilon$-values. This is not the case when moving horizontally in the direction of $V_0$, as the difference between the FI curves grows further apart for increasing $\epsilon$-values (see \ref{fig:FHNm_effect_V0_w0_global}, box to the right).\par

% How do the parameters correspond to the gs-derivatives
To translate the effect of parameters $V_0$ and $w_0$ to the measurable quantity of $g_s(V_{th})$, we look at the effect of the partial derivatives of $g_s(V_{th})$ (see eqn. \ref{eq:gs}). In the mFHN model, the saddle should be on the non-zero slope part of the $w$-nullcline, which corresponds to  $\frac{\partial{w_{\infty}}}{\partial{V}} \neq 0$. Figure  \ref{fig:gs_derivatives_local_global} shows that $\frac{\partial{w_{\infty}}}{\partial{V}}$ has the same trajectory as $w_0$, and affects the FI curve similarly, i.e. $I_{rheo}$ is affected (see fig.  \ref{fig:gs_derivatives_local_global} `local') and moving down the $\frac{\partial{w_{\infty}}}{\partial{V}}$-path increases the spike frequency (see fig. \ref{fig:gs_derivatives_local_global} `global'). The path of $\frac{\partial{\dot{V}}}{\partial{w}}$ reflects how the membrane voltage is affected by the slow gating variable. This effect is prominent around the transcritical bifurcation (white line), as shown in the parameter plots of figure \ref{fig:gs_derivatives_local_global}. In addition, $\frac{\partial{\dot{V}}}{\partial{w}}$ has a similar effect as $V_0$, where locally the $I_{rheo}$ is the same, whereas globally the frequency changes and the distance between FI curves increases for increasing $\epsilon$-values. The combined effect of  $\frac{\partial{\dot{V}}}{\partial{w}}$ and $\frac{\partial{w_{\infty}}}{\partial{V}}$ explains the effect seen in $g_s(V_{th})$, and allows to study the behaviour of $g_s(V_{th})$ and its partial derivatives around the pitchfork bifurcation (yellow star in parameter plots of fig. \ref{fig:gs_derivatives_local_global}). A $g_s(V_{th})$-path has similar values for $g_s(V_{th})$, but at different locations in the parameter plot. Comparing similar $g_s(V_{th})$-values close at the pitchfork bifurcation, shows that these values are more affected by movement in the horizontal direction, i.e. by $\frac{\partial{\dot{V}}}{\partial{w}}$, then the vertical direction, i.e.  $\frac{\partial{w_{\infty}}}{\partial{V}}$. This is also visible in the FI curves around the pitchfork bifurcation, as the FI curves share similar $I_{rheo}$ values, but differs globally in frequency, reflecting the large contribution of  $\frac{\partial{\dot{V}}}{\partial{w}}$. Continuing in the $g_s(V_{th})$-paths, $\frac{\partial{w_{\infty}}}{\partial{V}}$ is growing in influence and reflects the movement of $I_{rheo}$, while the global frequency remains similar. Overall, to obtain a similar parameter combination for Type I excitability around the pitchfork bifurcation, we show that $\frac{\partial{\dot{V}}}{\partial{w}}$ allows for a similar tuning as $V_0$ and $\frac{\partial{w_{\infty}}}{\partial{V}}$ reflects $w_0$.

% Bistability
An additional effect that characterises Type I excitability around the pitchfork bifurcation is the bistability that occurs in the region below the transcritical bifurcation line (white line in parameter plots of fig. \ref{fig:gs_derivatives_local_global}), which is detected using two different protocols for obtaining the FI curves, i.e. step-up and step-down protocol \cite{drion2015ion}. The last column of figure \ref{fig:gs_derivatives_local_global} shows a small bistability region for $\epsilon < \epsilon_{crit}$, however, for $\epsilon > \epsilon_{crit}$ a clear bistability region is present for locations close to the pitchfork bifurcation. For points further away from the pitchfork bifurcation the bistability is small. This effect is observed for paths in both the $g_s(V_{th})$-direction and the $\frac{\partial{\dot{V}}}{\partial{w}}$-direction, whereas paths in $\frac{\partial{w_{\infty}}}{\partial{V}}$ show a small change in bistability. \par

\afterpage{
	\clearpage% To flush out all floats, might not be what you want
	\begin{landscape}
		\thispagestyle{lscape}
		\pagestyle{lscape}
		\begin{figure}
			\includegraphics[width=22cm]{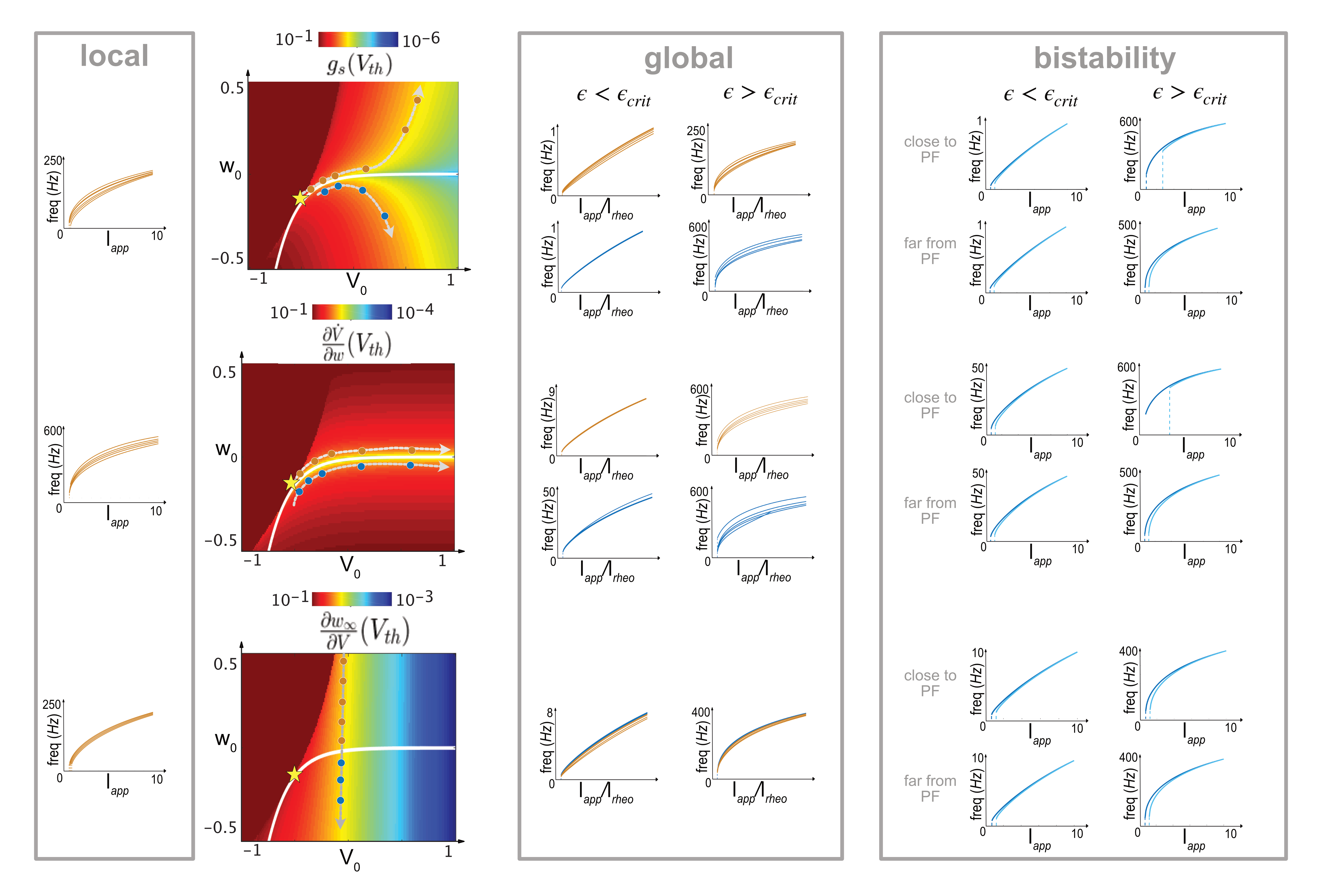}
			\caption{\textbf{The local, global, and bistability effects of $g_s$ and its derivatives on the frequency current curve.}  The first row shows the effect of $g_s$, the middle and last insets are the effect of the $g_s$ partial differential equations $\frac{\partial{V}}{\partial{w}}$ and $\frac{\partial{w_{\infty}}}{\partial{V}}$ respectively. The first column shows the \textit{local} effects, of the paths as indicated in the parameter space. The paths are directions for which a similar quantitative value is obtained at different locations in the parameter space. The local effects indicate the similarity of  $\frac{\partial{V}}{\partial{w}}$ to the effects of $V_0$ and $\frac{\partial{w_{\infty}}}{\partial{V}}$ to $w_0$.
			The second column shows the \textit{parameter space} with coordinates that follow paths above the transcritical bifurcation line, i.e. orange dots, and below the transcritical bifurcation line, i.e. blue dots. For all plots, orange lines correspond to the orange dots, which are the paths above the transcritical bifurcation line, and blue lines correspond to the blue dots which indicate the paths below the transcritical bifurcation line. The \textit{global} effects of the partial differentials are depicted in the third column, and indicates the global similarities of $\frac{\partial{V}}{\partial{w}}$ and $\frac{\partial{w_{\infty}}}{\partial{V}}$ to $V_0$ and $w_0$, respectively. The last column shows the effect of $g_s$ and its derivatives on the trajectories below the TC line for which \textit{bistability} occurs. The dark blue lines are the FI curves created following a step-up protocol and light blue lines are created using a step-down protocol. Here, a difference is observed for coordinates close to the pitchfork bifurcation (yellow star) and points further away from the pitchfork bifurcation. Parameter plot: yellow star: pitchfork bifurcation, white line: transcritical bifurcation; Statistics of FI curves are in Supplementary Material.}
			\label{fig:gs_derivatives_local_global}
		\end{figure}
	\end{landscape}
}

%%%%%%%%%%%%%%%%%%%%%%%%%%%%%%%%%%%%%%%%%%%%%%%%%%%
%%%%%%%%%%%%%%%%% 			Discussion						%%%%%%%%%%%%%%%%
%%%%%%%%%%%%%%%%%%%%%%%%%%%%%%%%%%%%%%%%%%%%%%%%%%%
\section{Discussion}
Phenomenological models are key in understanding dynamical behaviour of a neuronal excitable system. They reflect physiological observations and are a generalisation of the dynamics obtained from CBMs. This allows models, such as FHN and mFHN, to make predictions about neuronal behaviour and highlights the parameters that can be implemented in neuronal networks. Recent developments showed that slow gating channels can be divided into having restorative and regenerative properties, which is generalised in the mFHN model \cite{franci2013balance}. In this work, we show that the classification of Type I and Type II excitability solely based on the underlying bifurcation is incomplete, and needs the incorporation of a slow regenerative parameter that allows for zero-frequency onset in Type I excitability. This result is not only a mathematical condition, but is also found in reduced CBM models and is translatable to experimentally measurable values via the $\epsilon_{crit} = g_s(V_{th})$ relationship. In addition, we show that the coordinates in the mFHN model that indicate the region around the pitchfork bifurcation are translatable to the partial derivatives that are able to tune $g_s(V_{th})$ in an experimental biological setting to obtain Type I excitability. \par

%______________________		 gs = ecrit	_______________________	%
In mFHN, the incorporation of the slow regenerative gating results in the relationship $\epsilon_{crit} = g_s(V_{th})$, meaning that the change in bifurcation can be experimentally obtained by measuring the slow ion conductance at the voltage threshold, $V_{th}$. This relationship is also found in reduced CBM, where $g_s(V_{th}) = \frac{C}{\tau_n}\Bigr\rvert_{V_{th}} $, and underlines the generality of the mFHN model. The importance of this finding is that an actual physiological quantitative measure can be given to the situation where the system changes from one type of excitability to another type. During experiments, a physiologist can actually measure the slow conductance at the membrane voltage threshold and predict whether the neuron changes its state. \par

%____________________	Continuum Type I and II, and how to obtain Type I	___________________	%
Although neurons tend to be divided in either Type I excitability or Type II excitability, most neurons operate somewhere in between, such as neocortical pyramidal cells \cite{tateno2004threshold}. This continuum is also reflected in the FI curves around the $\epsilon_{crit}$-value. Here, the FI curves are indistinguishable, although they underlie a different type of bifurcation. Looking at the FI curves only would not indicate a change in excitability, but rather questions whether we should continue relating a SNIC bifurcation to a Type I excitability. In addition, the typical narrative for Type I excitability is that the trajectory is slowed down by a `ghost' region upon SNIC bifurcation. The mFHN model shows that the creation of a $V$-nullcline funnel, before the trajectory goes through the `ghost' region,  ensures the creation of an infinite period oscillation. Therefore, a SNIC bifurcation is not the decisive feature that underlies Type I bifurcation, but the existence of a region where the trajectory is slowed down to infinite period. In addition, the time-scale separation, $\tau$, should be large enough (small $\epsilon$) to go through the funnel, indicating that the recovery variable should be slow enough to create Type I excitability. Decreasing the slow recovery process to get Type I excitability was found before in the HH model and this insight has led to the experimental confirmation that increasing the temperature in the squid giant axon, decreases the slow recovery variable, and results in a Type I effect \cite{rinzel2013nonlinear}. \par

%______________________		 Parameters that make up FI	_______________________	%
The region in the mFHN where it possible to obtain Type I excitability is around the pitchfork bifurcation, which correspond to a low value for  $g_s(V_{th})$ in addition to a small transmembrane depolarizing current during ISI. Therefore, it is important to understand how the coordinates $V_0$ and $w_0$ in the mFHN model correspond to the partial derivatives that make up $g_s(V_{th})$. We found that the $g_s$ partial differentials, $\frac{\partial{\dot{V}}}{\partial{w}}$ and $\frac{\partial{w_{\infty}}}{\partial{V}}$, have the same effect as the parameters $V_0$ and $w_0$, respectively. The Type I excitability region is also characterised by a clearly distinguishable bistability effect. Overall, this shows that Type I excitability that corresponds to the region around the pitchfork bifurcation in the mFHN model reflects the behaviour of partial derivatives that constitute $g_s(V_{th})$.\par

%______________________		 Future 	_______________________	%
Together, these results show that further generalisation of the mFHN model by incorporating slow regenerative current, reflects observations seen in physiology, such as Type I excitability and the continuous change between Type I and Type II excitability. These results in spike initiation dynamics have a direct influence on neuronal input-output properties in network coding, such as regulating synchrony transfer and network coding strategies \cite{ratte2013impact}. To translate between CBMs and network models, an integrate-and-fire (IF) model can be used that is based on the parameter values, such as the adaptive exponential IF model \cite{brette2005adaptive} or Multi-Quadratic Integrate-and-Fire (MQIF) model, which is a generalisation of the classical Quadratic Integrate-and-Fire (QIF) model \cite{vanPottelberg2018robust}. The MQIF model is based on the time-scale generalisation as described by the DIC, and incorporates the bistability caused by the slow regenerative conductance. The importance of incorporating time-scale separation in spike initiation dynamics into network models is beneficial to understand neuromodulation and homeostasis \cite{gjorgjieva2016homeostatic}, \cite{oleary2014cell}, as well as circuit organisation involved in synaptic plasticity during learning and memory \cite{gjorgjieva2016computational}, \cite{gjorgjieva2011triplet}.

%%%%%%%%%%%%%%%%%%%%%%%%%%%%%%%%%%%%%%%%%%%%%%%%%%%
%%%%%%%%%%%%%%%%			Methods				%%%%%%%%%%%%%%%%%%%%%
%%%%%%%%%%%%%%%%%%%%%%%%%%%%%%%%%%%%%%%%%%%%%%%%%%%

\section{Methods} 
	
	%________________________		 Reduced CS	_________________ %
	\subsection{Reduced Connor-Stevens model}
	The Connor-Stevens model is reduced following the methods described in \cite{franci2014modeling} and \cite{drion2015ion}. To recite: sodium channel activation and A-type potassium channel activation are merged in the fast timescale [$m_{Na} = m_{Na,\infty}(V)$ and $m_A = m_{A,\infty}(V)$]. Delayed-rectifier potassium channel activation, sodium channel inactivation, and A-type potassium channel inactivation variables are merged into a single slow variable $n$ [$m_{Kd} = n$, $h_{Na} = h_{Na,\infty}(n_{\infty}^{-1}(n))$, $h_A = h_{A,\infty}(n_{\infty}^{-1}(n))$]. We set $n_{\infty}(V) \equiv m_{Kd, \infty}(V)$. Because $m_{Kd, \infty}(V)$ is not invertible in closed form in the original CS model, we use the exponential fit (see \cite{drion2015ion}), which results in 
	
	\begin{equation*}
	m_{Kd, \infty}^{-1} = - (200 \log(1/V - 21/20))/13 - 208/5.
	\end{equation*}
	
	\noindent
	The system of equations for the reduced CS:
	
	\begin{alignat*}{2}
	&C\dot{V} = & &-g_{Na} m_{\infty}(V)^3 h_{\infty}(n_{\infty}^{-1}(n))(V-V_{Na}) \\
	& & &-g_K n^4(V-V_K) -g_A m_{A,\infty}(V)^3 h_{A, \infty}(n_{\infty}^{-1}(n))(V-V_K) \\
	& & &-g_L(V-V_L) + I_{app} \\
	&\dot{n} = & & 1/\tau_n(V) (n_\infty(V_0 - V) - n)\\	
	\end{alignat*}	
	
	\noindent
	The steady-state activation curves are approximated by the Boltzmann function, which has the form
	
	\begin{equation*}
	x_{\infty}(V) = \frac{1}{1 + \exp(V_0 - V)/k)},
	\end{equation*}
	
	\noindent
	where $V_0$ is the $V_{half}$. The following procedure describes how to get from the $\alpha,\beta$ form, $x_{\infty}(V) = \frac{\alpha_x(V)}{\alpha_x(V) + \beta_x(V)}$ to the Boltzmann function.
	
	\begin{enumerate}
		\item Get $x_{\infty}(V_0) = 1/2$ using the $\alpha,\beta$ equation. 
		
		\item Get the slope at $V_0$, i.e. $V_{slope}(V_0)$, by differentiating the $\alpha,\beta$ equation and plug in the value obtained in the previous step. 
		%Then use the general point-slope equation $y = V_{0,slope}([V_{start}:V_{end}] - V_0) + 1/2$ to get the slope equation. 
		
		\item Get $k$ as described in \cite{izhikevich2007dynamical} pg. 45. 
		
		\begin{equation*}
		k = \frac{V_0 - V_{slope}(V_0)}{2}
		\end{equation*}
		
	\end{enumerate}

	%________________________		 Reduced HH	_________________ %
	\subsection{Reduced Hodgkin-Huxley with Ca$^+$model}
	The Hodgkin-Huxley with Ca$^+$ was reduced as described in \cite{drion2012novel}. The standard reduction of the original HH was followed with the additional assumption of the correlation between the potassium and calcium gating kinetics being $d := n^3$, where $d$ is the calcium activation gating variable and $n$ the potassium gating variable.

	The system of equations is
	\begin{alignat*}{2}
	&C\dot{V} = & &-g_{Na} m_{\infty}(V)^3 (0.89 - 1.1 *n)(V-V_{Na}) \\
	& & &-g_K n^4(V-V_K) -g_{Ca} n^3 (V-V_{Ca}) \\
	& & &-g_L(V-V_L) + I_{pump} + I_{app} \\
	&\dot{n} = & & 1/\tau_n(V) (n_\infty(V_0 - V) - n)\\	
	\end{alignat*}	
	
	The steady-state activation curves for $m_{\infty}$ and $n_{\infty}$ are approximated as described in the reduced CS section.

	%________________________		 Calculate TC	_________________ %
	\subsection{Calculation transcritical bifurcation}
	To calculate the transcritical bifurcation we have 3 parameters that are unknown, $V_0, w_0, I_{app}$. Therefore, the following three conditions need to be met (see \cite{franci2013balance}):
	
	\begin{enumerate}
		\item singularity equation for bifurcation, which is $det(J) = 0$. 
		
		\begin{align*}
		\frac{\partial\dot{V}}{\partial{V}} \frac{\partial\dot{w}}{\partial{w}} - \frac{\partial\dot{V}}{\partial{w}} \frac{w_{\infty}}{\partial{V}} &= 0 \\
		\frac{\partial\dot{V}}{\partial{V}} (-1) - \frac{\partial\dot{V}}{\partial{w}} \frac{w_{\infty}}{\partial{V}} &= 0 \\
		\frac{\partial\dot{V}}{\partial{V}} + \frac{\partial\dot{V}}{\partial{w}} \frac{w_{\infty}}{\partial{V}} &= 0
		\end{align*}
		
		\item the slow conductance $g_s = 0$, i.e. 
		
		\begin{equation}
		\frac{\partial\dot{V}}{\partial{w}} \frac{\partial{w_{\infty}}}{\partial{V}} = 0
		\end{equation}
		
		\item and the $V$-nullcline intersect with itself, i.e.
		
		\begin{equation}
		\frac{\partial\dot{V}}{\partial{V}} = 0
		\end{equation}
		
	\end{enumerate}

	\subsubsection{Transcritical bifurcation in mFHN} 
	In the mFHN, $I_{app}(TC) = 2/3$ and $V_{TC} = -1$, see \cite{franci2013balance}. Therefore, only $w_0(TC)$ needs to be calculated. This is done by solving for $\dot{V} = 0$ and $\dot{w} = 0$.
	
	\begin{align*}
	&w_0(TC) =  0 \\
	&V_{TC} - \frac{V_{TC}^3}{3} - \left( \frac{2}{(1+e^(-5*(V_{TC}-V_0)) + w_0)^2} \right) + I_{app}(TC) =  0
	\end{align*}

	\subsubsection{Transcritical bifurcation in reduced CS}
	In the reduced CS, $g_A$ is the slow conductance, corresponding to $w_0$ in the mFHN. We obtain $g_A(TC)$ using condition (3.). Recall the reduced CS model:
	
	\begin{alignat*}{2}
	&C\dot{V} = & &-g_{Na} m_{\infty}(V)^3 h_{\infty}(n_{\infty}^{-1}(n))(V-V_{Na}) \\
	& & &-g_K n^4(V-V_K) -g_A m_{A,\infty}(V)^3 h_{A, \infty}(n_{\infty}^{-1}(n))(V-V_K) \\
	& & &-g_L(V-V_L) + I_{app} \\
	&\dot{n} = & & \frac{1}{\tau_n(V)} (n_\infty(V) - n)\\	
	\end{alignat*}	
	
	The procedure is as follows:
	\begin{enumerate}
		\item We extract the $g_A$ using condition (3.), i.e. $\frac{\partial{\dot{V}}}{\partial{V}} = 0$. We get
		
		\begin{alignat*}{2}
		&\frac{\partial{\dot{V}}}{\partial{V}} = & & \Big( -g_{Na} \frac{\partial}{\partial{V}} m_{\infty}(V)^3 h_{\infty}(n_{\infty}^{-1}(n))(V-V_{Na})\\
		& & & + (-g_{Na} m_{\infty}(V)^3 h_{\infty}(n_{\infty}^{-1}(n)) \frac{\partial}{\partial{V}} (V-V_{Na}) )\\
		& & &-g_K n^4 \frac{\partial}{\partial{V}} (V-V_K) \\
		& & & -g_A \frac{\partial}{\partial{V}} m_{A,\infty}(V)^3 h_{A, \infty}(n_{\infty}^{-1}(n))(V-V_K) \\
		& & & + (-g_A m_{A,\infty}(V)^3 h_{A, \infty}(n_{\infty}^{-1}(n))\frac{\partial}{\partial{V}} (V-V_K))\\
		& & &-g_L\frac{\partial}{\partial{V}}(V-V_L) \Big) = 0 \\
		\end{alignat*}
		
		\begin{alignat*}{2}
		&\frac{\partial{\dot{V}}}{\partial{V}} = & &  -g_{Na} \left[ 3m_{\infty}(V)^2 h_{\infty}(n_{\infty}^{-1}(n))(V-V_{Na})+ m_{\infty}(V)^3 \right] \\
		& & & -g_K n^4 \\
		& & & -g_A \left[ 3m_{A,\infty}(V)^2 h_{A, \infty}(n_{\infty}^{-1}(n))(V-V_K) + m_{A,\infty}(V)^3 \right]\\
		& & & -g_L.
		\end{alignat*}
		
		And thus $g_A$ becomes
		
		\begin{equation}
		g_A = \dfrac{\splitdfrac{
				-g_{Na} \left[ 3m_{\infty}(V)^2 h_{\infty}(n_{\infty}^{-1}(n))(V-V_{Na})+ m_{\infty}(V)^3 \right]}{ -g_K n^4 -g_L }}{3m_{A,\infty}(V)^2 h_{A, \infty}(n_{\infty}^{-1}(n))(V-V_K) + m_{A,\infty}(V)^3}	
		\end{equation}

		\item The $V_{TC}$ is obtained using condition (2.) and the obtained expression for $g_A$. 
		
		\item Then $n_{TC}$ is obtained using $n_\infty$, i.e. 
		
		\begin{equation*}
		n_{\infty} = \frac{1}{1 + \exp((V_0 - V_{TC})/k_n)}
		\end{equation*}
		
		\item Now we can get the $g_A$ at transcritical bifurcation using condition (3.) and solve with $\frac{\partial\dot{V}}{\partial{V}}|_{(V_{TC}, n_{TC}, g_A)}$
		
		\item Finally, $I_{app}(TC)$ is obtained using the singularity condition (1.)
		
		\begin{equation*}
		\begin{split}
		I_{app}(TC) = \\
		&\quad -( (-g_{Na} m_{\infty}(V_{TC})^3 h_{\infty}(n_{\infty}^{-1}(n_{TC}))(V_{TC}-E_{Na})\\
		&\quad -g_{KDR} n_{TC}^4(V_{TC}-E_K) \\
		&\quad -g_{A, TC} m_{A,\infty}(V_{TC})^3 h_{A, \infty}(n_{\infty}^{-1}(n_{TC}))(V_{TC}-E_K)\\
		&\quad -g_L(V_{TC}-E_L)) )
		\end{split}
		\end{equation*}
		
	\end{enumerate}

	\subsection{Numerical analysis}
	Parameter space plots, phase planes and linearity of $\epsilon_{crit} = g_s(V_{TH})$ (figs. \ref{fig:bfnChange}, \ref{fig:FHNnc_epsilon_crit}, \ref{fig:mFHN_epsilon_crit}, \ref{fig:gsecrit}, \ref{fig:FHNm_effect_V0_w0_local}, \ref{fig:FHNm_effect_V0_w0_global}, \ref{fig:gs_derivatives_local_global}, \ref{fig:CBMred_bifn_map}) were produced using MATLAB (available at http://www.mathworks.com). The simulations, FI curves and stochastic simulations (figs. \ref{fig:bfnChange}, \ref{fig:FHNnc_epsilon_crit}, \ref{fig:FHNm_effect_V0_w0_local}, \ref{fig:FHNm_effect_V0_w0_global}, \ref{fig:gs_derivatives_local_global}) were performed in Julia (available at https://julialang.org/). All code created is archived under \href{https://zenodo.org/badge/latestdoi/113469222}{DOI: https://zenodo.org/badge/latestdoi/113469222}
	The TD-diagram in fig. \ref{fig:bfnChange} were hand-drawn using Affinity Designer vector graphics editor (https://affinity.serif.com/en-us/).

%%%%%%%%%%%%%%%%%%%%%%%%%%%%%%%%%%%%%%%%%%%%%%%%%%%%
%%%%%%%%%%%%%			Acknowledgements		%%%%%%%%%%%%%%%%%%%%%
%%%%%%%%%%%%%%%%%%%%%%%%%%%%%%%%%%%%%%%%%%%%%%%%%%%%
\section{Acknowledgements}
The authors wish to acknowledge the insightful comments and suggestions by Dr. Alessio Franci.

%%%%%%%%%%%%%%%%%%%%%%%%%%%%%%%%%%%%%%%%%%%%%%%%%%%
%%%%%%%%%%%%%%			Supplementary material			%%%%%%%%%%%%%%%%%
%%%%%%%%%%%%%%%%%%%%%%%%%%%%%%%%%%%%%%%%%%%%%%%%%%%
\section{Supplementary Material}

%________________________		 Oscillation		_________________ %
\subsection{Oscillation}

\begin{figure}[H]
	\centering
	\includegraphics[width=1.0\linewidth]{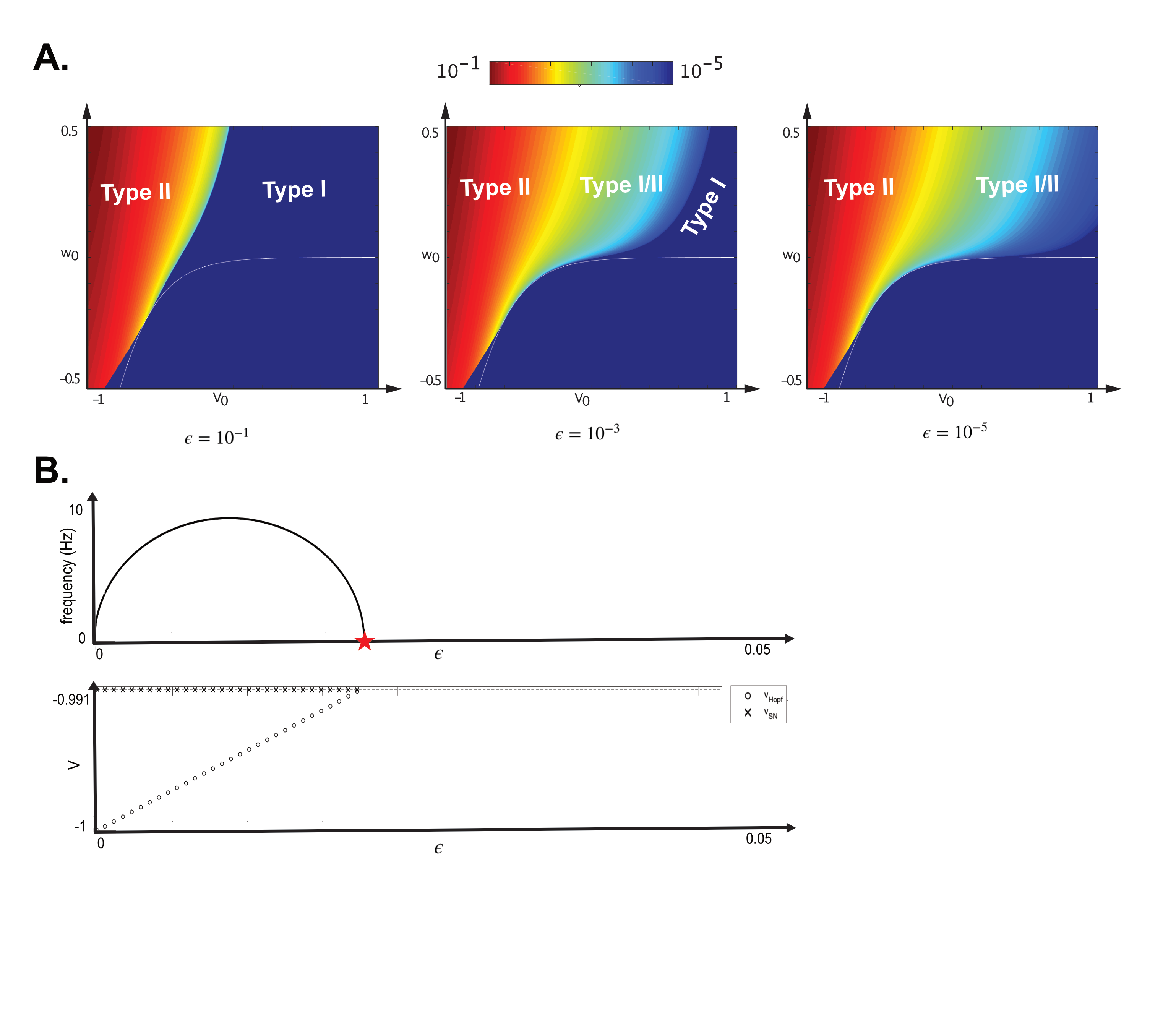}
	\caption{\textbf{Oscillation frequency of left fixed point}. The bifurcation maps in \textbf{A} plot the imaginary part of the left fixed point eigenvalues. The $Im(\lambda)$ reflects the oscillation frequency of the left fixed point as $\omega = Im(\lambda)$. $\omega$ is exact at birth of the limit cycle (see \cite{strogatz2018nonlinear}) and its frequency is visualised with the colour scale. When the system has a small time scale separation, i.e. $\epsilon = 10^{-1}$, the region next to Type II is almost completely Type I. Although in these region the system contains three fixed point, the $\epsilon$-value is such that the left fixed point is stable in almost all cases and SNIC bifurcation occurs. Decreasing the $\epsilon$-value increases the time-scale separation and the situation for which the left fixed point is destabilised via subcritical Hopf. Typically, this results in the existence of a small oscillation cycle around the left fixed point. In the case of large time scale separation, $\epsilon = 10^{-5}$, oscillation is visible almost throughout the entire Type I/II region. In addition, the frequency is large just after the Hopf bifurcation area but gradually decreases when $V_0$ increases. \textbf{B} shows the frequency of the small unstable oscillation cycle that appears upon subcritical Hopf bifurcation. The horizontal axis is the change in $\epsilon$. The frequency decreases to zero at the Bogdanov-Takens (BT) bifurcation (red star). (see \cite{izhikevich2007dynamical}, pg. 196) The lower plot indicates the location of the saddle point and the destabilised left fixed point.}
	\label{fig:FHNm_Hopffreq_eps}
\end{figure}

Upon destabilisation of the left fixed point by subcritical Hopf bifurcation, a small unstable limit (oscillation) cycle is created. The imaginary part of the eigenvalues of the left fixed point indicates the size of the limit cycle (\cite{strogatz2018nonlinear}). The resulting frequency of this limit cycle indicates the subthreshold oscillation seen in neurons before spiking. As Hopf bifurcation occurs in a system with one fixed point as well as in the situation with three fixed point where the slow conductance is restorative, it is expected that the amount of oscillation in the system is dependent on the $\epsilon$-value. Figure \ref{fig:FHNm_Hopffreq_eps} shows that a large $\epsilon$-value ($\epsilon = 10^{-1}$) results in a system with a clear division between Type II in a system with one fixed point, and Type I in a system with three fixed points. There is just a very small Type I/II region in the system with three fixed points. This region is growing when $\epsilon$-decreases is decreased, i.e. where the time-scale coupling factor $\tau$ increases. At $\epsilon = 10^{-5}$, the region Type I/II has expanded throughout the system with three fixed points. The gradual change of oscillation as observed in figure \ref{fig:TC_Hopffreq_eps} shows that the oscillation frequency is decreasing when moving from the `Type II' region. This indicates that the timescale coupling factor not only reflects the oscillation in the system, but also gradually changes the system from a strictly divided 'Type II' and 'Type I' region to a 'Type II' and 'Type I/II' region. \par

%________________________		 Bogdanov Takens		_________________ %
\subsection{Boganov Takens}
At the $\epsilon_{crit}$-value, the equilibrium undergoes Hopf and SNIC bifurcation simultaneously, i.e. Hopf=SNIC. Therefore, the Jacobian matrix satisfies $det(J)=0$ and $tr(J)=0$, which are the conditions for Bogdanov-Takens (BT) bifurcation (\cite{izhikevich2007dynamical}). During a BT bifurcation the SNIC and Hopf are accompanied by a saddle homoclinic orbit (SHO) bifurcation. This SHO occurs when left fixed point undergoes the subcritical Hopf bifurcation upon which a small unstable oscillation cycle is created. This cycle degenerates into an homoclinic orbit to the saddle and disappears via a SHO bifurcation (\cite{izhikevich2007dynamical}, pg 196). Figure \ref{fig:FHNm_Hopffreq_eps} \textbf{B.} indicates the frequency of the small orbit cycle over a range of $\epsilon$-values and shows that this frequency decreases to zero at the BT bifurcation (red star). As shown in the lower plot of figure \ref{fig:FHNm_Hopffreq_eps} \textbf{B.}, the destabilised left fixed point moves closer to the saddle point when the $\epsilon$-value approaches the $\epsilon_{crit}$-value. Therefore, at $\epsilon_{crit}$-value a subcritical Hopf bifurcation, a SNIC bifurcation and a SHO bifurcation occurs, which is characteristic for the BT bifurcation. \par

%_______________________			 SN-SH 		_______________________	%
\subsection{SN-SH} In the regenerative region, below the transcritical bifurcation line in the bifurcation map, the saddle node lies on the lower branch of the $V$-nullcline. In this situation, a saddle-node saddle-homoclinic (SN-SH) bifurcation underlies a Type II* (\cite{drion2015ion}) or Type IV (\cite{franci2012organizing}) excitability. The SN-SH bifurcation exhibit a bistable range, which is in contrast to the Type I and II excitabilities. The bistability occurs when $\epsilon < \epsilon_{crit}$, as the vector field is such that the trajectory is either on limit cycle or stopped by stable left fixed point (see fig. \ref{fig:mFHN_epsilon_crit}). When  $\epsilon > \epsilon_{crit}$, all trajectories end at the stable left fixed point and SNIC bifurcation is necessary for spiking to occur.

%________________________		 FI curves further from TC line		_________________ %
\subsection{FI curves further from transcritical bifurcation line}

\begin{figure}
	\centering
	\includegraphics[width=0.7\linewidth]{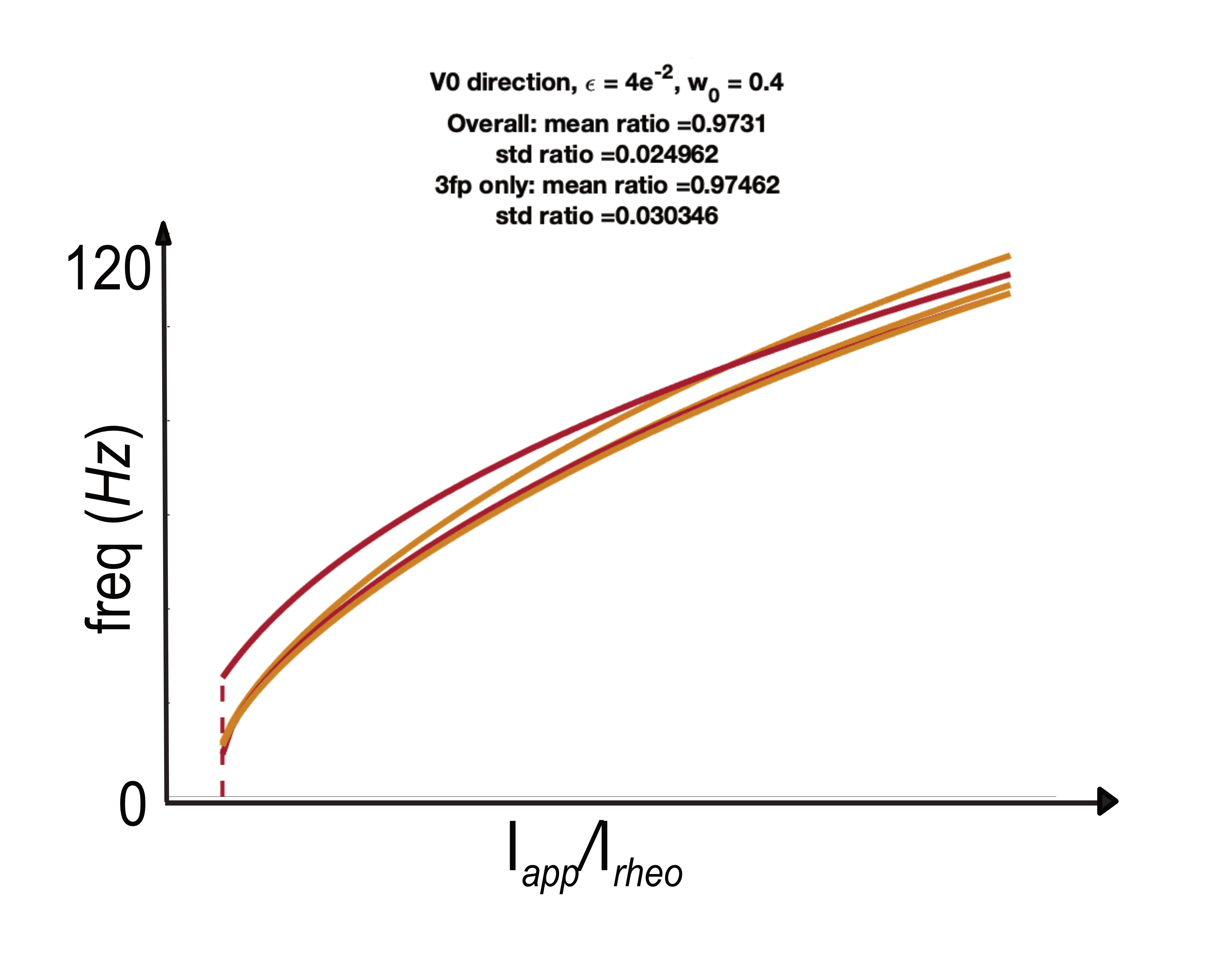}
	\caption{\textbf{Global effect of parameter $V_0$ further from transcritical bifurcation line.} This is in addition to figure \ref{fig:FHNm_effect_V0_w0_global} where points are shown that are close around the transcritical bifurcation lline. Red curve is an FI curve from the Hopf only region, i.e. Hopf in a system with one fixed point. Overall statistics are for all FI curves, 3fp statistics are for FI curves in three fixed point system (orange curves)}
	\label{fig:FHNm_effect_V0w0_supplementary}
\end{figure}

Figure \ref{fig:FHNm_effect_V0_w0_global} showed the effect of $V_0$ and $w_0$ on the FI curves. The coordinates shown are close to the transcritical bifurcation line. The effect of parameters on the FI curve is also indicated for points further away from the transcritical bifurcation line, as shown in figure \ref{fig:FHNm_effect_V0w0_supplementary}. The statistics are indicated above the plot.

%_______________________			 FI curve statistics 		_______________________	%
\subsection{FI curve statistics}
The following tables describe the statistics of the FI curves in figure \ref{fig:FHNm_effect_V0_w0_global}. The table below depicts the sample mean $\bar{X}$ and the standard deviation (std) of $V_0$ above the transcritical (TC) bifurcation and the $w_0$ FI curves.

% V0 above TC and w0 
\begin{tabular}{ccccc}
	\hline
	\multirow{2}{*}{} & \multicolumn{2}{c}{$V_0$ above TC} & \multicolumn{2}{c}{$w_0$} \\
	& $\bar{X}$ 																& std & $\bar{X}$ & std \\
	\hline
	$\epsilon < \epsilon_{crit}$ 										& 0.99374 & 0.0060154 & 0.96452 & 0.011255 \\
	$\epsilon_{{crit}_L} < \epsilon < \epsilon_{{crit}_R}$ & 0.98157 & 0.016973 & 0.97292 & 0.0089168\\
	$\epsilon > \epsilon_{crit}$ 										& 0.94182 & 0.043362 & 0.9884 & 0.0048987 \\
	\hline
\end{tabular}

\vskip 10pt
\noindent
The next table describes the same statistics for $V_0$ below the transcritical bifurcation line.

% V0 below TC 
\begin{tabular}{ccc}
	\hline
	\multirow{2}{*}{} & \multicolumn{2}{c}{$V_0$ below TC} \\
	& $\bar{X}$ 																& std \\
	\hline
	$\epsilon < \epsilon_{crit}$ 										& 0.99664 &  0.004011 \\
	$\epsilon_{{crit}_L} < \epsilon < \epsilon_{{crit}_R}$ & 0.98071 & 0.015171 \\
	$\epsilon > \epsilon_{crit}$ 										& 0.94541 & 0.053218  \\
	\hline
\end{tabular}

\vskip 10pt
\noindent
The following tables describe the statistics for the FIs that follow paths in the $g_s$-direction and in the $g_s$ derivatives direction of figure \ref{fig:gs_derivatives_local_global}. The first table describes the sample mean and std for $g_s$ paths above and below the transcritical bifurcation line.

% gs above and below TC
\begin{tabular}{ccccc}
	\hline
	\multirow{2}{*}{} & \multicolumn{2}{c}{$g_s$ above TC} & \multicolumn{2}{c}{$g_s$ below TC} \\
	& $\bar{X}$ 																& std & $\bar{X}$ & std \\
	\hline
	$\epsilon < \epsilon_{crit}$ 										& 0.9694 & 0.022506 & 0.99178 & 0.0061999 \\
	$\epsilon > \epsilon_{crit}$ 										& 0.96501 & 0.019762 & 0.94461 & 0.017761 \\
	\hline
\end{tabular}

\vskip 10pt
\noindent
The following table describes the statistics for the FIs that follow the paths in $\frac{\partial \dot{V}}{\partial w}$.

% dvdw above and below TC
\begin{tabular}{ccccc}
	\hline
	\multirow{2}{*}{} & \multicolumn{2}{c}{$\frac{\partial \dot{V}}{\partial w}$ above TC} & \multicolumn{2}{c}{$\frac{\partial \dot{V}}{\partial w}$ below TC} \\
	& $\bar{X}$ 																& std & $\bar{X}$ & std \\
	\hline
	$\epsilon < \epsilon_{crit}$ 										& 0.9936 & 0.006375 & 0.98249 & 0.022646 \\
	$\epsilon > \epsilon_{crit}$ 										& 0.95187 & 0.012489 & 0.94023 & 0.036778 \\
	\hline
\end{tabular}

\vskip 10pt
\noindent
This last table are the FI stats for $\frac{\partial {w_{\infty}}}{\partial V}$ paths

% dwinfdv all 
\begin{tabular}{ccc}
	\hline
	\multirow{2}{*}{} & \multicolumn{2}{c}{$\frac{\partial {w_{\infty}}}{\partial V}$} \\
	& $\bar{X}$ 																& std \\
	\hline
	$\epsilon < \epsilon_{crit}$ 										& 0.97174 &  0.017012 \\
	$\epsilon > \epsilon_{crit}$ 										& 0.9875 & 0.0078541  \\
	\hline
\end{tabular}

%_______________________			 Noise 		_______________________	%
\vskip 40pt
\subsection{Effect of noise on Type I excitability}
Besides Hodgkin's classification of excitability, neurons can be further categorised depending on their response to input noise  \cite{lundstrom2008two}, \cite{chance2002abbott}, \cite{phillips2008alpha}, \cite{arsiero2007impact}. This neuronal sensitivity to input fluctuations are thought to provide an additional communication channel for neural coding \cite{fairhall2001efficiency} and promote persistent activity in networks \cite{arsiero2007impact}. In \cite{lundstrom2009sensitivity} is shown that the separation of time-scale influences the sensitivity to input noise. 
Adding white noise to the input, as described in \cite{goldwyn2011and}, shows that for $\epsilon < \epsilon_{crit}$ the frequency remains stable and the coefficient of variation (CV) is low, which is not the case for $\epsilon > \epsilon_{crit}$ (fig. \ref{fig:gs_derivatives_local_global}, box `noise'). The reason for the increased variability when $\epsilon$ is large is due to the more oblique vector in the phase plane. Adding noise causes more difficulty to go through the funnel created by $V$-nullcline (see fig. \ref{fig:overview_models_TDdiagram}). Choosing a location close around pitchfork bifurcation by adjusting $\frac{\partial{\dot{V}}}{\partial{w}}$ and $\frac{\partial{w_{\infty}}}{\partial{V}}$ causes the trajectory to not go through the funnel with the added noise. As the noise added is based on the overall highest and lowest $\epsilon_{crit}$-values of the coordinates, this means that the $\epsilon$-value is high for points around the pitchfork bifurcation. Therefore, it seems that there is an upper boundary for $\epsilon$ for which the vector is too oblique to create a spiking behaviour around the pitchfork bifurcation, and thus Type I excitability is only possible when the $\epsilon$-value is small enough, i.e. the time-scale coupling $\tau_w$ is large. \par

\begin{figure}[h!]
	\centering
	\includegraphics[width=0.6\linewidth]{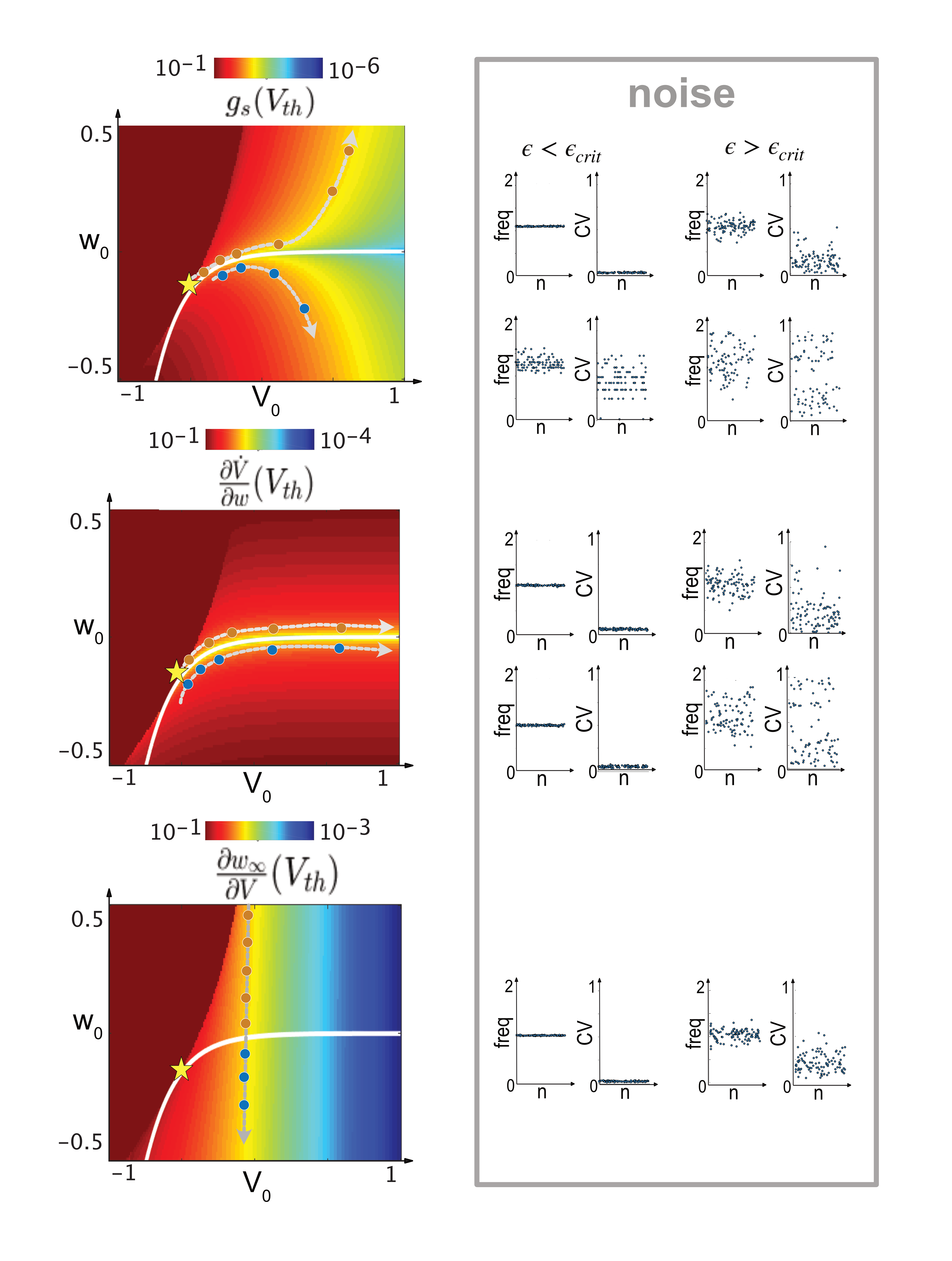}
	\caption{\textbf{Effect of noise on $g_s$ and its derivatives.} The effect of white noise on the input is shown in the fourth column. The frequency is normalised for the nominal frequency, i.e. the frequency without added noise. The coefficient of variantion (CV) shows a similar effect as the normalised frequency. Parameter plot: yellow star: pitchfork bifurcation, white line: transcritical bifurcation; Parameter: white noise $\sigma = 0.5$, simulations $n = 100$.}
	\label{fig:gsderivatives_noise}
\end{figure}

%%%%%%%%%%%%%%%%%%%%%%%%%%%%%%%%%%%%%%%%%%%%%%%%%%%
%%%%%%%%%%%%%%%%%		Bibliography			%%%%%%%%%%%%%%%%%%%%
%%%%%%%%%%%%%%%%%%%%%%%%%%%%%%%%%%%%%%%%%%%%%%%%%%%
\newpage
\bibliographystyle{unsrt}  
\bibliography{../../ref.bib}

\end{document}